\documentclass[aps,prd,onecolumn,groupedaddress,nofootinbib]{revtex4}         

\usepackage[english]{babel}

\usepackage[letterpaper,top=2cm,bottom=2cm,left=3cm,right=3cm,marginparwidth=1.75cm]{geometry}

\usepackage{amsmath}
\usepackage{graphicx}
\usepackage[colorlinks=true, allcolors=blue]{hyperref}

\begin{document}

\title{Cosmic-Dawn 21-cm Signal from Dynamical Dark Energy
}

\author{Lu Yin$^{1}$}
\email{lu.yin@apctp.org}
\affiliation{$^{1}$Asia Pacific Center for Theoretical Physics, Pohang, 37673, Korea}
%\affiliation{$^{2}$Department of Physics, POSTECH, Pohang 37673, Korea}

\begin{abstract}
The 21-cm signal is the most important measurement for us to understand physics during cosmic dawn. 
It is the key for us to understand the expansion history of the Universe and the nature of dark energy.
In this paper, we focused on the characteristic 21-cm power spectrum of a special dynamic dark energy - the Interacting Chevallier-Polarski-Linder (ICPL) model - and compared it with those of the $\Lambda$CDM and CPL models. From the expected noise of HERA, we found more precise experiments in the future can detect the features of interacting dark energy in the 21-cm power spectra. By studying the brightness temperature, we found the ICPL model is closer to the observation of EDGES compared to the $\Lambda$CDM, thus alleviating the tension between theory and experiments.

%Chevallier-Polarski-Linder (CPL) model 
%local star-formation rate density (SFRD)
\end{abstract}

\maketitle

\section{Introduction}

The Experiment to Detect the Global Epoch of Reionization Signature (EDGES) has reported the detection of an absorption profile centered at {78 MHz} \cite{Bowman:2018yin}, corresponding to a redshift $z \approx 17$, which is named the global 21-cm signal. 
The 21-cm signal refers to the wavelength of the electromagnetic radiation emitted by neutral hydrogen atoms  at cosmic dawn. Specifically, it corresponds to the transition between the two hyperfine energy levels of the hydrogen atom, where the electron's spin and the proton's spin are aligned (parallel) and anti-aligned (anti-parallel) with each other \cite{Pritchard:2011xb, Morales:2009gs, Furlanetto:2006jb}.
The global 21-cm signal  provides valuable information about the distribution and properties of neutral hydrogen gas in the Universe. It is particularly significant for studying the epoch of reionization, the formation of the first galaxies and stars, and the large-scale structure of the Universe \cite{Tashiro:2014tsa, Feng:2018rje, Barkana:2018qrx, Mahdawi:2018euy, Mirocha:2018cih, Ewall-Wice:2018bzf, Hirano:2018alc, Venumadhav:2018uwn, Clark:2018ghm, Munoz:2018jwq, Pospelov:2018kdh, Li:2018kzs, Hektor:2018lec, Li:2018okf}.
However, the amplitude of the absorption profile reported by EDGES is more than twice the maximum allowed in the standard cosmological evolution, which implies the new physics beyond the standard model \cite{Bowman:2018yin}.

To generate a possible strong signal in the 21-cm line, two mechanisms are commonly considered. The first mechanism involves enhancing the background radiation temperature through processes such as dark matter decay or annihilation \cite{Fraser:2018acy, Yang:2018gjd, DAmico:2018sxd, Mitridate:2018iag, Cheung:2018vww}. These models propose that the presence of additional energy injected into the system leads to an increase in the background radiation temperature, which in turn affects the 21-cm signal. The second mechanism focuses on lowering the gas temperature through interactions between dark matter and baryons \cite{Barkana:2018lgd, Kovetz:2018zan, Slatyer:2018aqg, Munoz:2018pzp, Berlin:2018sjs, Barkana:2018qrx}. By considering elastic scattering or other interactions between these particles, the thermal state of the gas can be modified, resulting in a different 21-cm signal.
However, many existing theories face challenges when confronted with other astronomical observations. 
%They are inconsistent with other data sets or theoretical predictions. 
People are looking forward to a model involving additional cooling or heating mechanisms to solve this tension. 

The interactions between dark matter and dark energy can influence the gas temperature and, consequently, the 21-cm signal \cite{Xiao:2018jyl, Costa:2018aoy, Wang:2018azy, Li:2019loh}.
%The interacting dark energy (IDE) model  proposes the energy transformation between dark energy and dark matter. 
On the one hand, interacting dark energy (IDE) can be fitted with other observations well and can give different 21-cm predictions. On the other hand, the 21-cm signal will help to constrain the equation of state of dark energy and distinguish dark energy models. In this work, we consider the phenomenology of 21-cm global signal for two special dynamical dark energy models: the Chevallier-Polarski-Linder (CPL) model \cite{Chevallier:2000qy, Linder:2002et, Colgain:2021pmf} and the Interacting CPL (ICPL) model \cite{He:2008tn}. We will compare the difference in the 21-cm global signal with the $\Lambda$CDM by using the star-formation rate density (SFRD) method from the $Zeus21$ program \cite{Munoz:2023kkg}. 
%The calculation using the star-formation rate density (SFRD) method from the $Zeus21$ program will save much time in numerical calculation \cite{Munoz:2023kkg}. 
%signal-to-noise detection
%Additionally, the modification of the background evolution has been considered as a potential explanation for the observed anomalous 21-cm signal. Models such as Early Dark Energy (EDE) and Interacting Dark Energy (IDE) propose modifications to the background evolution of the Universe, which can impact the thermal state of the gas and result in deviations from the standard 21-cm signal predictions.

%Overall, researchers have explored various mechanisms and scenarios to explain the observed deviations in the 21-cm line signal, considering interactions between dark matter and baryons, additional cooling or heating mechanisms, and modifications to the background evolution. These investigations aim to reconcile the observed data with our current understanding of the early Universe. 
%21 cm helps us to understand the evolution of the Universe, and build up the equation of the state of dark energy.
%21CM can help to distinguish DE models.

%For the 21-cm observatories, both global-signal efforts and interferometers, are dedicated to studying the 21-cm global signal from neutral hydrogen during the cosmic dawn and the epoch of reionization (EoR). These

Except EDGES, there are other 21-cm observations, such as the Shaped Antenna measurement of the background Radio Spectrum (SARAS)\cite{Singh:2017syr}, the Large-aperture Experiment to Detect the Dark Ages (LEDA) \cite{Peitzmann:1996xv}, the Low-Frequency Array (LOFAR)\cite{vanHaarlem:2013dsa}, the Murchison Widefield Array (MWA)\cite{Beardsley:2016njr, Tingay:2012ps}, the Hydrogen Epoch of Reionization Array (HERA)\cite{DeBoer:2016tnn, HERA:2021noe}, and the upcoming Square Kilometre Array (SKA) \cite{Santos:2015gra, Zhang:2019dyq, Xu:2020uws, Wang:2022oou, Hartley:2023ach}. We will  also compare our theoretical model predictions with the HERA data by imploring the expected noise.

%expected noise from HERA

The paper will proceed as follows. We review the calculation of 21-cm brightness temperature in Section \ref{sec:2}. And Section \ref{sec:3} shows the background evolution of the CPL, the ICPL, and the $\Lambda$CDM model with their best-fit results in local observations. In Section \ref{sec:4}, we calculate the 21-cm brightness temperature signals from the three models. Moreover, we compare the power spectra of the three models with the  expected noise from HERA.
The conclusion is presented in Section \ref{sec:5}. 
%we review the Boltzmann equations and how they are modified by the presence of an explicit coupling between a scalar field and photons. In Section \ref{sec:3} we review the early dark energy models that we explore. Our main results are presented in Section \ref{sec:4}; the EB power spectra and the tools we use to discriminate between different models. We conclude in Section \ref{sec:5}. 

\section{21-CM LINE BRIGHTNESS TEMPERATURE}
\label{sec:2}
The 21-cm line is produced when the electron and proton in a hydrogen atom are in opposite spin states, which is known as hyperfine splitting. When the electron and proton flip their spins and move to the other spin state, they emit or absorb a photon with a frequency of about {1420 MHz}, corresponding to a wavelength of 21 cm. 
The brightness temperature $T_{21}$ is defined as the temperature an observer would measure from a 21-cm line observation and describes the strength of the global sky-average signal, which can be given by:
\begin{equation}
T_{21}=\frac{T_{S}-T_{\mathrm{CMB}}}{1+z}\left(1-e^{-\tau_{21}}\right),
\end{equation}
where $T_S$ is the spin temperature of the hydrogen atom, $T_{\mathrm{CMB}}$ is the temperature of the cosmic microwave background radiation, and $z$ is the redshift of the observed 21-cm line. The spin temperature is a measure of the population difference between the two hyperfine energy levels of the hydrogen atom, and it depends on the collisional and radiative processes.
%that can change the level populations. Thus, the brightness temperature $T_{21}$ is an important quantity for studying the properties of the intergalactic medium and the early universe.

The $\tau_{21}$ represents the 21-cm optical depth of the diffuse intergalactic medium \cite{Barkana:2000fd} and can be calculated as
\begin{equation}
\label{eq:optical}
\tau_{21}=(1+\delta) x_{\mathrm{HI}} \frac{T_{0}}{T_{S}} \frac{H(z)}{\partial_{r} v_{r}}(1+z),
\end{equation}
where $x_{\mathrm{HI}}$ is the fraction of neutral hydrogen, $\delta$ is the density of the neutral hydrogens, $H(z)$ is the Hubble parameter, and $\partial_{r} v_{r}$ is the velocity line-of-sight gradient. The normalization factor $T_0$ can be defined by
\begin{equation}
T_{0}=34 \mathrm{mK} \times\left(\frac{1+z}{16}\right)^{1 / 2}\left(\frac{\Omega_{b} h^{2}}{0.022}\right)\left(\frac{\Omega_{m} h^{2}}{0.14}\right)^{-1 / 2},
\end{equation}
where the $\Omega_m$ and $\Omega_b$ are the fractions of the critical density for cold dark matter and baryonic matter.
Another important parameter is the{
color temperature $T_c$, which is closely associated with the gas kinetic temperature $T_k$  \cite{Barkana:2004vb} and numerically approximated by
%$T_k$ is the gas kinetic temperature.
\begin{equation}
T_{c}\approx T_k^{-1}+g_{col} T_k^{-1}(T_{S}^{-1}-T_k^{-1}),
\end{equation}
where $g_{col}\approx 0.4055 $ $K$. 
By taking the approximation $\tau_{21}\ll 1$
%, and linear-order redshift-space distortions 
\cite{Barkana:2005jr, Mao:2011xp}, we can re-write the 21-cm brightness temperature
\begin{equation}
T_{21}=T_0(z)(1+\delta-\delta_v)(\frac{x_\alpha}{1+x_\alpha})(1-\frac{T_{\mathrm{CMB}}}{T_c})x_{HI},
\label{eq:5}
\end{equation}
where $x_\alpha$ is the dimensionless Wonthuysen-Field (WF) coupling parameter \cite{Munoz:2023kkg, Hirata:2005mz}.} 
{Thus, the equation for the 21-cm signal, $T_{21}$, separates into four different parts. 
These are:
(i) $(1+\delta-\delta_v)$ represents the large-scale structure, which is described by the local density and velocity fields. 
(ii) $(x_\alpha/(1+x_\alpha))$ is the term of the WF effect, which measures the Lyman-alpha emission from the first galaxies.
(iii) The term $(1-T_{\mathrm{CMB}}/T_c)$ can be considered by the gas kinetic and color temperature, which is determined by the competition between adiabatic cooling and X-ray heating due to the first stars.
(iv) Reionization is represented by the term $x_{HI}$.}
By considering the four terms, the 21-cm signal can be used as a powerful tracer of the formation of the first stars and galaxies, as well as the evolution of the cosmic web at high redshifts.
%The evolution of the 21-cm signal is closely linked to the intensity of the Lyman-alpha (through $x_\alpha$) and X-ray (through $T_c$) backgrounds emitted by the first galaxies. These radiative fields can travel significant distances before being absorbed, and their fluxes at a given point depend on the emission over the past lightcone.}

For cosmic dawn, the recombination process had already finished, and the neutral hydrogen number density had already been set. Therefore, the brightness temperature can be modified via a different background evolution. 
%The dominant effects on the 21-cm brightness temperature are due to the affected by the evolution of Hubble parameter.
%A modified Hubble parameter can change the redshift at which the coupling between the hydrogen atoms and the CMB radiation becomes efficient, and can also affect the rate at which the spin temperature equilibrates with the gas temperature. 
The dynamical dark energy models will provide a different evolution of Hubble parameter in our Universe and change the $T_{21}$.

%$$
%\begin{aligned}
%T_{21}(z) & \approx \frac{T_{S}-T_{C M B}}{1+z} \tau \\
%& \approx 0.023 \times x_{H I}\left(\frac{T_{S}-T_{C M B}}{T_{S}}\right)\left(\frac{\Omega_{b} h^{2}}{0.02}\right) \\
%& \times\left[\left(\frac{0.15}{\Omega_{m} h^{2}}\right)\left(\frac{1+z}{10}\right)\right]^{1 / 2} \mathrm{~K} .
%\end{aligned}
%$$
%$$
%\begin{aligned}
%\tau & =\frac{3 c^{3} \hbar A_{10} n_{H I}}{16 k_{B} v_{0}^{2} T_{S} H(z)} \\
%& \approx 8.6 \times 10^{-3} x_{H I}\left[\frac{T_{C M B}(z)}{T_{S}}\right]\left(\frac{\Omega_{b} h^{2}}{0.02}\right) \\
%& \times\left[\left(\frac{0.15}{\Omega_{m} h^{2}}\right)\left(\frac{1+z}{10}\right)\right]^{1 / 2} .
%\end{aligned}
%$$

%\begin{figure}
%	\centering
%\includegraphics[width=0.5\textwidth]{WCDM-1.pdf}
% \includegraphics[width=0.5\textwidth]{IWCDM-1.pdf}
%\includegraphics[width=0.483\textwidth]{lcdm-0418.pdf}
%\includegraphics[width=0.483\textwidth]{wcdm-0418.pdf}
%\includegraphics[width=0.483\textwidth]{iwcdm-0418.pdf}
%	\caption{\label{fig:theta}The evolution of scalar field with redshift.}
%\end{figure}

\section{Evolution of Dynamical Dark Energy Models}
\label{sec:3}
%We will introduce the background evolution of DDE models.
Based on the FRW Universe, each component of the Universe can be written in Friedmann equations as
\begin{equation}
3 M_p^2 H^2=\rho_r+\rho_b+\rho_{c}+\rho_{de},
\end{equation}
where the energy density of radiation, baryon, cold dark matter, and dark energy can be expressed as $\rho_r$, $\rho_b$, $\rho_{c}$, and $\rho_{de}$, respectively.
In the local Universe, dark energy and dark matter make up approximately $96\%$ components of the Universe. The change in their energy densities will lead to a different evolutionary history of our Universe.

%To produce the accurate expanding Universe,
%Dark energy can provide a negative pressure, which causes the accelerated expansion of the Universe.
The ratio between pressure and energy density is the equation of state ($\omega=P/\rho$).
In the $\Lambda$CDM, $\omega_{de}$ is considered as a constant number and equal to -1. But in dynamic dark energy, the equation of state can be evolved with redshift.
For example, in Chevallier-Polarski-Linder (CPL) model\cite{Chevallier:2000qy, Linder:2002et, Colgain:2021pmf}, the $\omega_{de}$ can be written as
\begin{equation}
\omega_{de}=\omega_0+\omega_a z /(1+z)=\omega_0+\omega_a(1-a).
\end{equation}
%where the $\omega_0$ and $\omega_a$ are constent numbers.
If $\omega_a$ returns to 0 and $\omega_0$ is -1, this model will go back to the $\Lambda$CDM.

%In the thermal history of cosmology, the recombination stage is related to 21-cm line signal. And this is also a matter-dominant area.  
The period of cosmic dawn is  matter dominated. 
%If any changes the background evolution during this time and change the thermal history and the 21-cm line signal.
The interacting dark energy model is one option to change the background evolution during this time and change the thermal history. 
We consider a special model that has energy transformation between dark energy and dark matter based on CPL, named the ICPL model. Their continuity equation for the energy densities can be parameterized as 
\begin{equation}
\begin{aligned}
\dot{\rho}_{c} & +3 H \rho_{c}=Q, \\
\dot{\rho}_{de} & +3 H\left(1+w_{de}\right) \rho_{de}=-Q,
\end{aligned}
\end{equation}
where $Q$ is the non-gravitational interacting energy transfer term.
%The different forms of $Q$  could lead to the different rates of this energy transform. 
There are already many studies on different forms of Q \cite{Cai:2004dk, Amendola:2006dg, He:2008tn, Li:2009zs, Li:2013bya, Geng:2017apd, Zhang:2018nga, Li:2019loh, Geng:2020mga, Geng:2020upf}.
In this study, we consider $Q$ as
\begin{equation}
Q=3\gamma H \rho_{c},
\label{eq:q}
\end{equation}
where the $\gamma$ is the interaction parameter, and we will fit it later. If the value of $\gamma$ comes back to zero, there is no interaction between dark energy and dark matter anymore, and the model will return to the CPL model. 
%With the non-zero value of $\gamma$, the model can be named as Interacting CPL model ($ICPL$). 
With the energy transfer as Eq.~(\ref{eq:q}), the evolution of cold dark matter and dark energy can be modified as
\begin{equation}
\rho_{c}=\rho_{c}^0(1+z)^{3(1-\gamma)},
\end{equation}

\begin{equation}
\begin{aligned}
\rho_{de}= & 3\gamma \rho_{c}^0 (1+z)^{3\left(1+\omega_0+\omega_a\right)} e^{3 \omega_a /(1+z)} \int_0^z e^{-3 \omega_a /(1+z)}(1+z)^{-3\left(\omega_0+\omega_a+\gamma \right)-1} d z \\
& +\rho_{de}^0(1+z)^{3\left(1+\omega_0+\omega_a\right)} e^{-3 \omega_a z/(1+z)}.
\label{eq:rhode}
\end{aligned}
\end{equation}

The definitions of the three models in background evolution are summarized in Tab.~\ref{tab:model}. 
\begin{table}
	\begin{center}
	\caption{{\color{black}The background evolution calculated from the  $\Lambda$CDM, the CPL, and the ICPL model, respectively.}}
	\begin{tabular}{|l|c|c|c|}
		 \hline $\text { Parameter } $& $\Lambda$CDM& CPL & ICPL \\
		\hline 
            $\omega(z)$      &$ -1 $ & $\omega_0+\omega_a z /(1+z) $      &$\omega_0+\omega_a z /(1+z) $\\
            \hline
		$Q$              &$ 0 $  & $ 0 $                 &$3\gamma H \rho_{c} $\\
  		\hline
		$\rho_{de}$      &$\rho_{de}^0 $   & $ \rho^0_{de}(1 + z)^{3(1+\omega_0+\omega_a)}e^{-\frac{3\omega_a z}{1+z}}$ & Eq.\ref{eq:rhode}\\
  		\hline
		$\rho_c $        &$ \rho_{c}^0(1+z)^{3}$                           & $ \rho_{c}^0(1+z)^{3} $   &$\rho_{c}^0(1+z)^{3(1-\gamma)}$\\
		\hline
		\end{tabular}
\label{tab:model}
\end{center}
\end{table}
Next, we will use the Markov Chain Monte Carlo (MCMC) to get the best-fit value of every parameter for the three models and discuss their effects on the 21-cm signal.

\section{Result and Discussion}
\label{sec:4}

\begin{table}
	\begin{center}
	\caption{{\color{black}The best-fit of free parameters comes from the $\Lambda$CDM,  the CPL, and the ICPL model, respectively. These results were obtained with 68$\%$ C.L. based on the observation of BAO and Type Ia Supernova.}}
	\begin{tabular}{|l|c|c|c|}
		 \hline $\text { Parameter } $& $\Lambda$CDM& CPL & ICPL \\
		\hline 
            $H_0 $      &$ 69.9182^{+0.1398}_{-0.1369} $ & $70.6986^{+0.2639}_{-0.2601} $      &$ 70.9338^{+0.2549}_{-0.2515}$\\
            		\hline
		$\Omega_b $ &$ 0.0445\pm 0.0004 $            & $ 0.0433\pm 0.0005 $          &$0.0440 \pm 0.0005 $\\
  		\hline
		$\Omega_{c}$&$0.2563^{+0.0075}_{-0.0071} $    & $ 0.2722^{+0.0100}_{-0.0096}$ &$ 0.2634^{+0.0106}_{-0.0101} $\\
  		\hline
		$w_0 $      &$ -1$                           & $-1.2132^{+0.0654}_{-0.0659} $   &$-1.0230^{+0.0692}_{-0.0626}$\\
  		\hline
		$w_a$       &$ 0 $                           & $0.6415^{+0.3945}_{-0.3822}$     &$0.6470^{+0.5976}_{-0.6253}$ \\
  		\hline
		$\gamma$    &$ 0 $                           & $ 0 $                          &$-0.0078^{+0.0192}_{-0.0118}$\\
		\hline
		\end{tabular}
\label{tab:2}
\end{center}
\end{table}

\begin{figure}
	\centering
\includegraphics[width=0.45\textwidth]{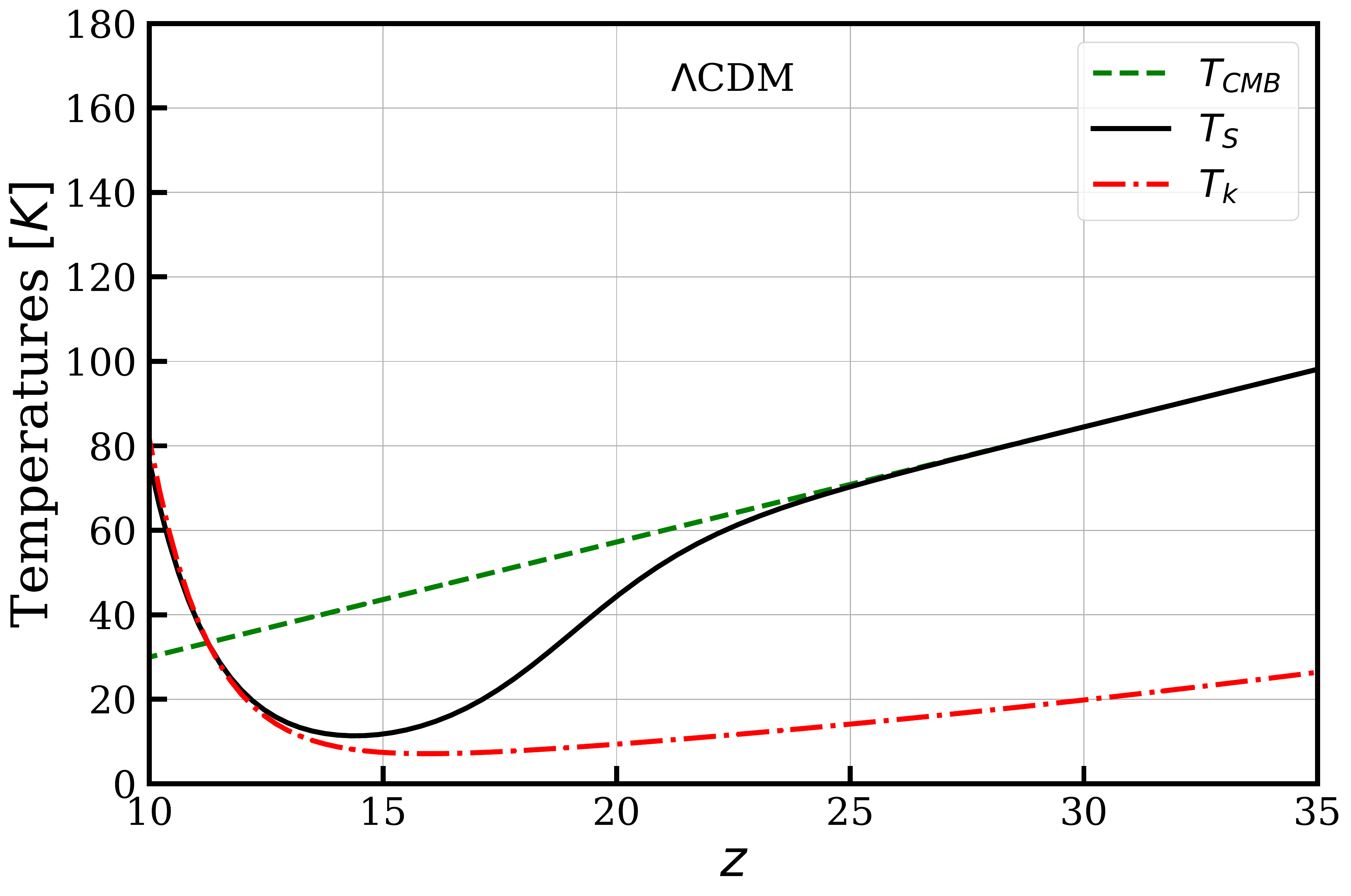}
 \includegraphics[width=0.45\textwidth]{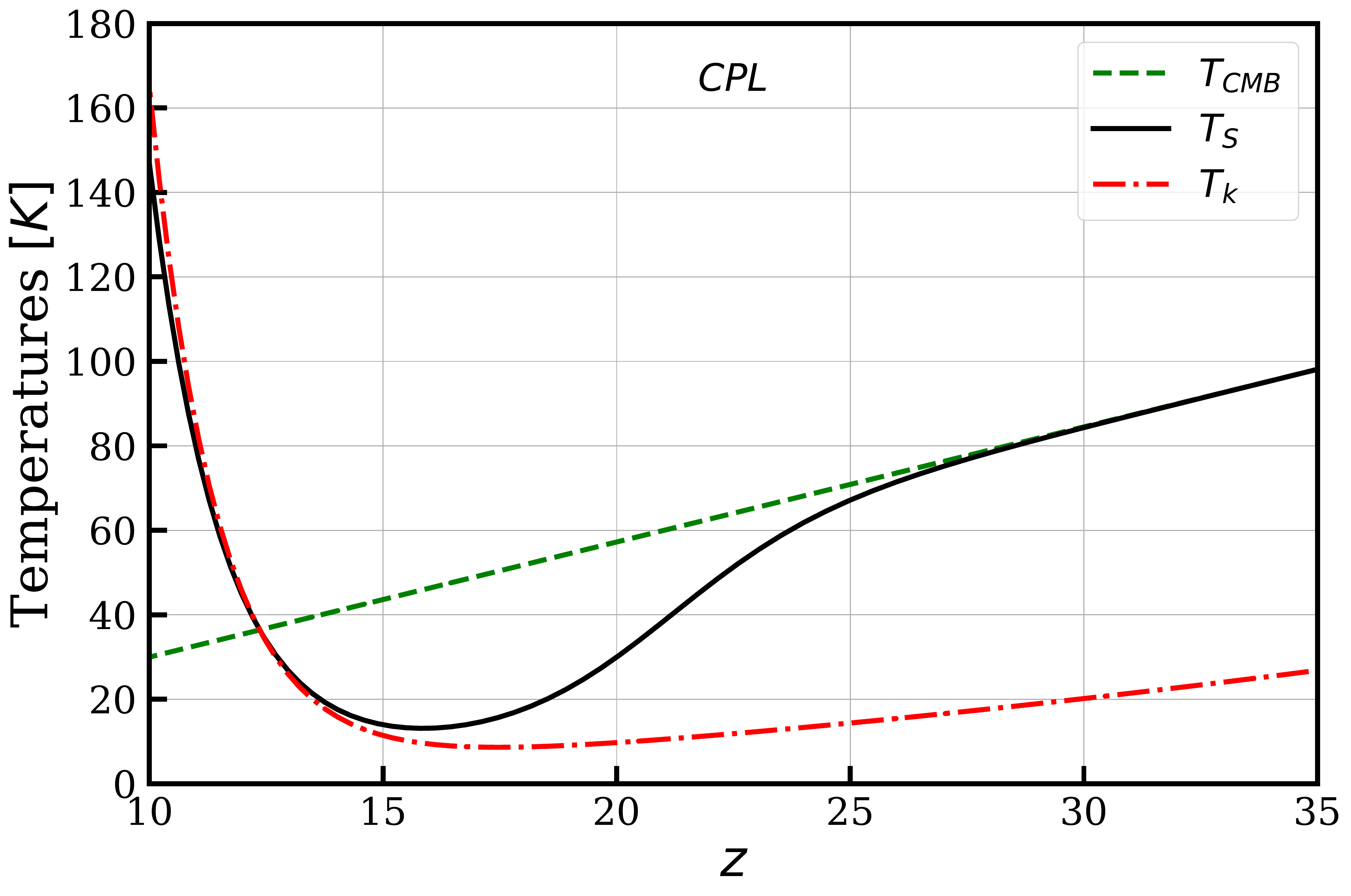}
 \includegraphics[width=0.45\textwidth]{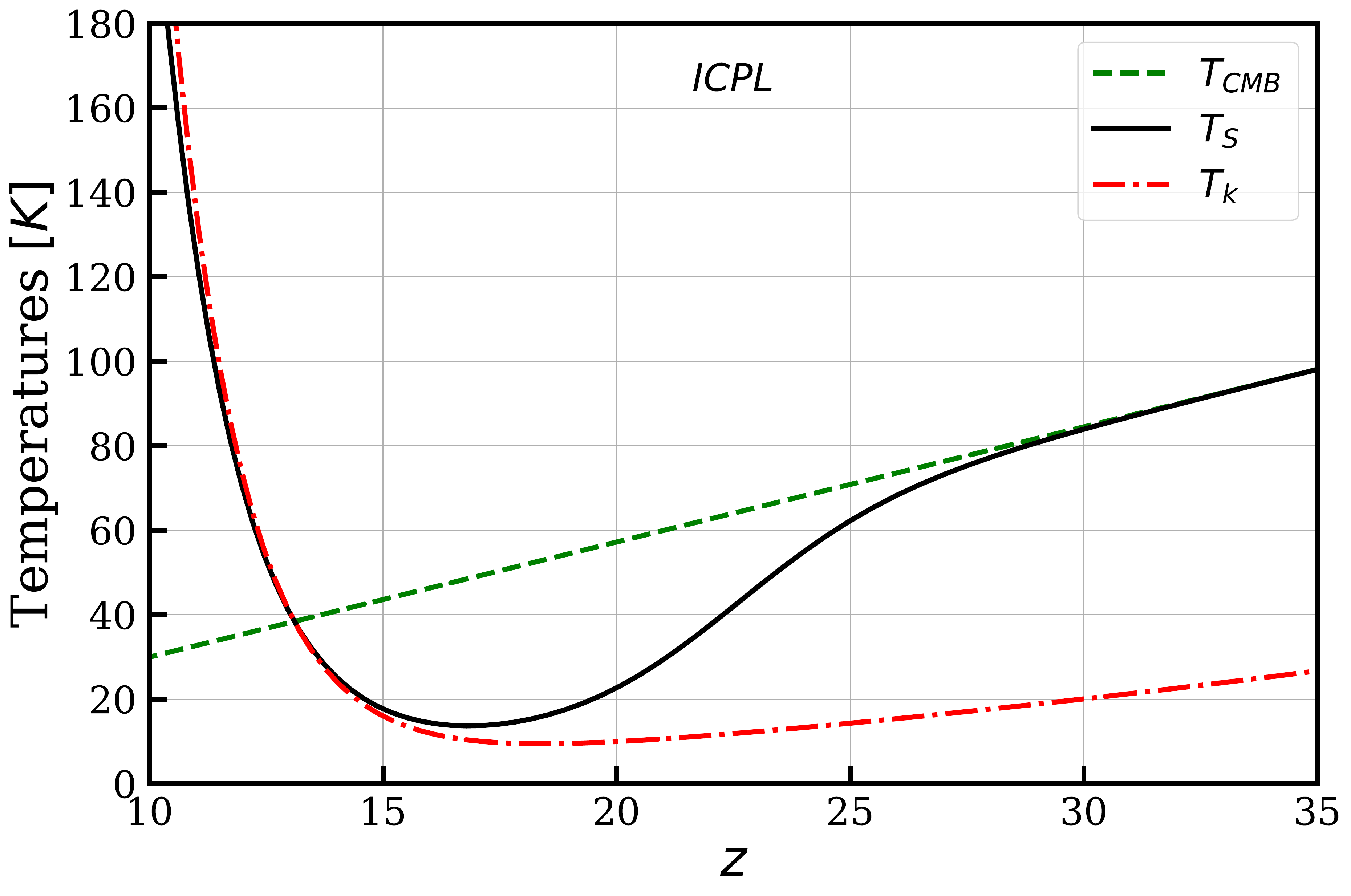}
	\caption{\label{fig:1}The evolution of the CMB temperature $T_{\mathrm{CMB}}$ (dash green line), spin temperature $T_S$ (black line), gas kinetic temperature $T_k$ (dash red line) with redshift in the $\Lambda$CDM, the CPL, and the ICPL model.}
\end{figure}

\begin{figure}
	\centering
 \includegraphics[width=0.45\textwidth]{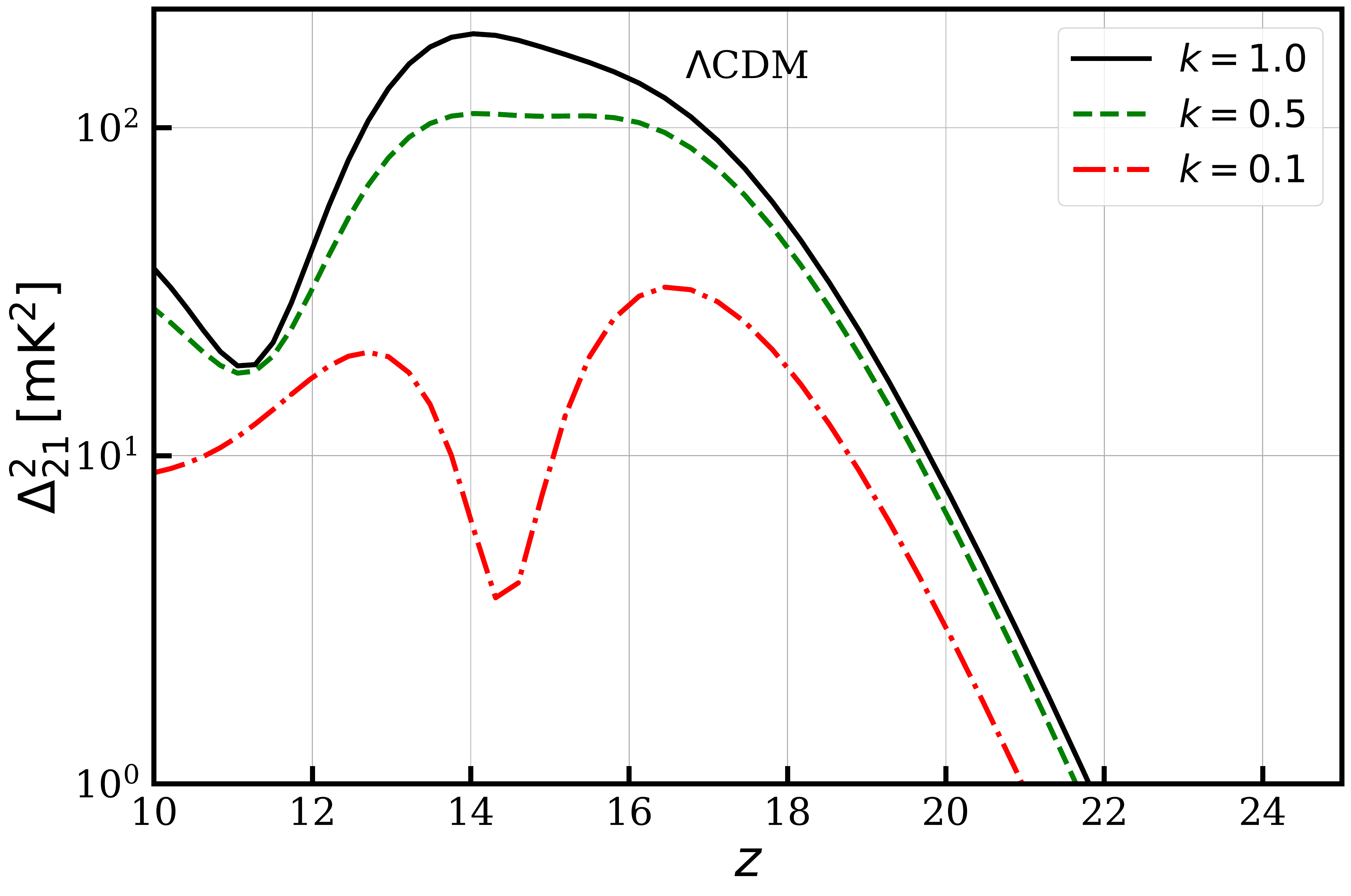}
 \includegraphics[width=0.45\textwidth]{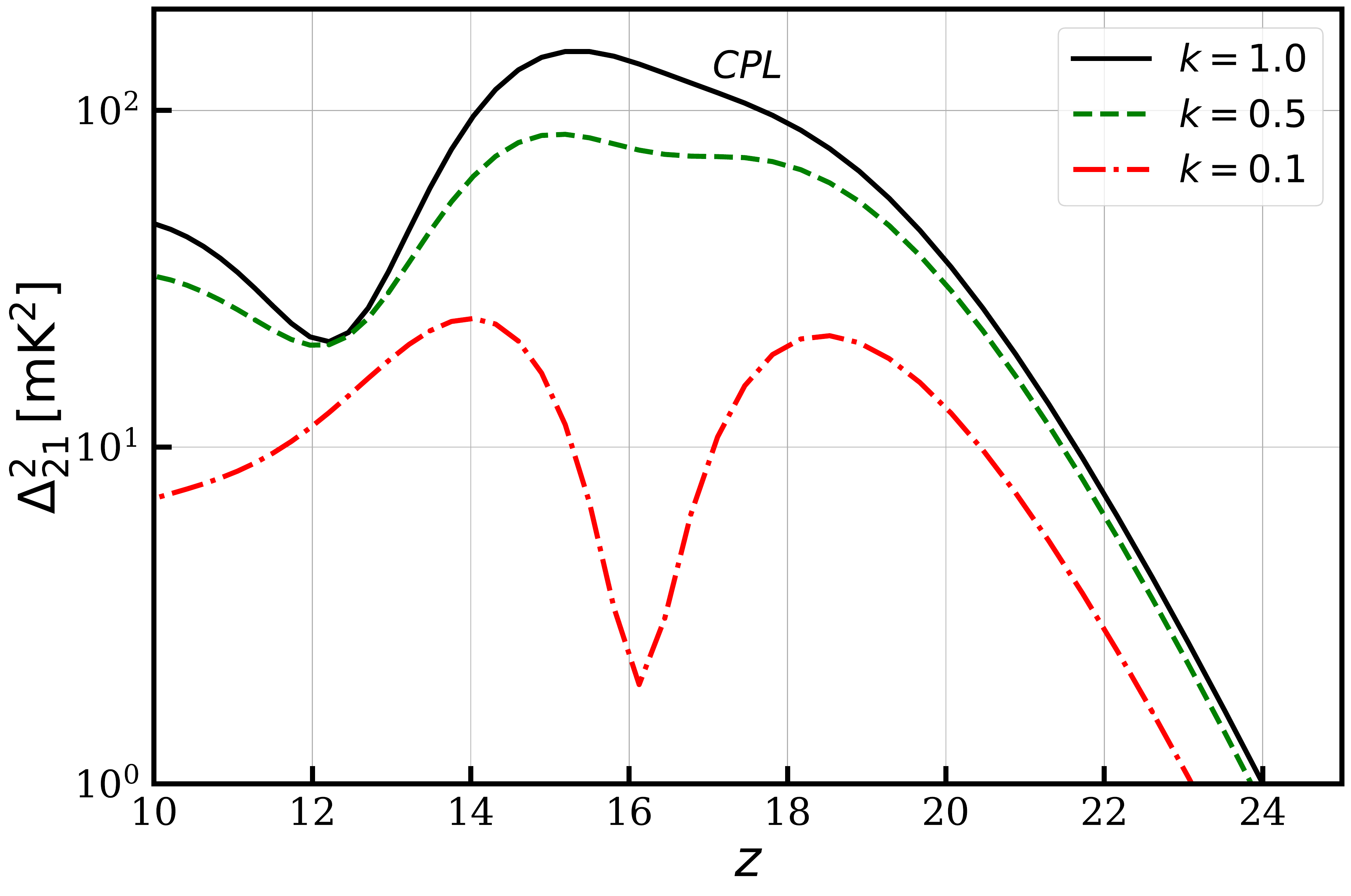}
   \includegraphics[width=0.45\textwidth]{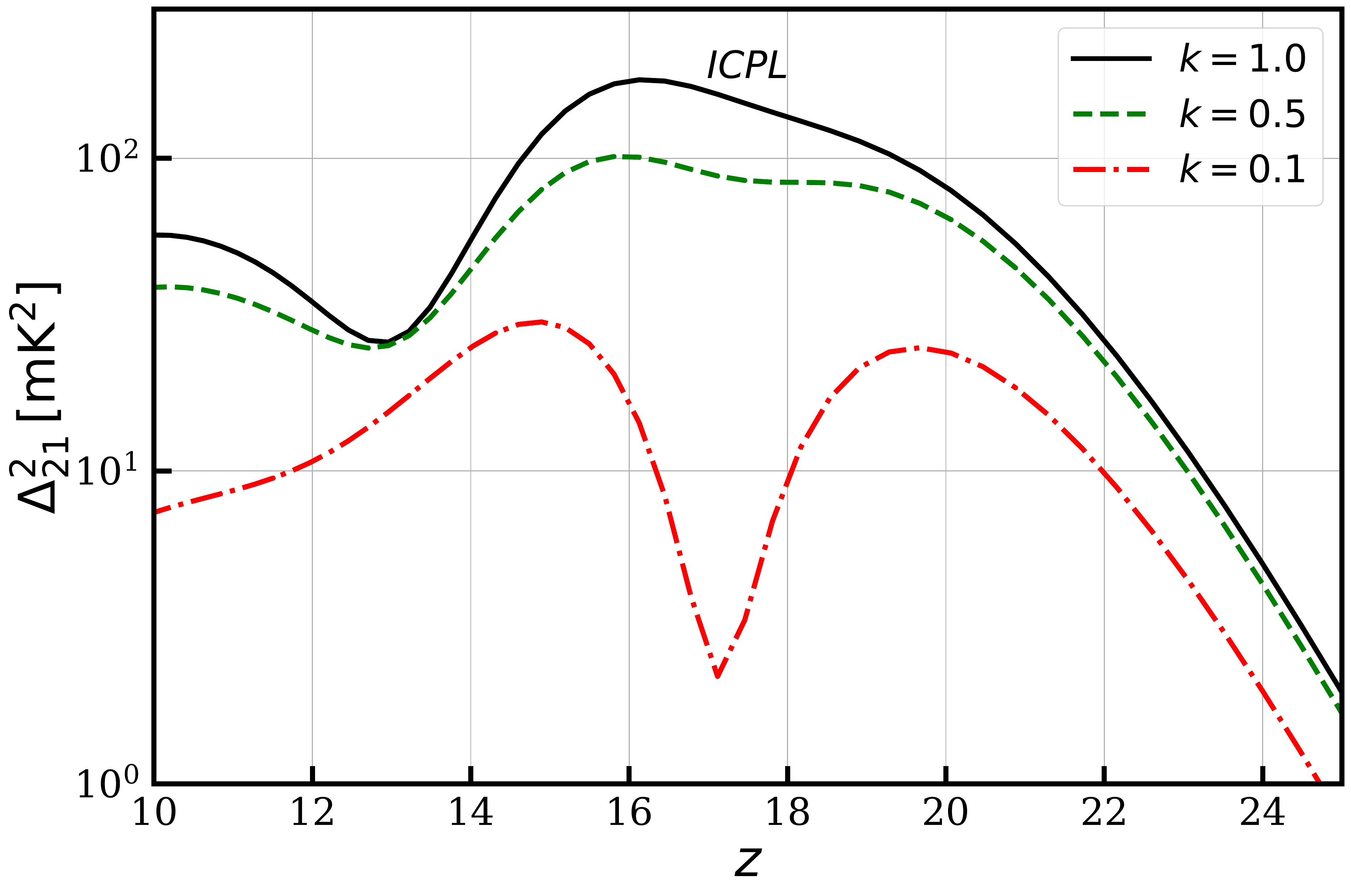}
	\caption{\label{fig:2}The evolution of 21-cm power spectra with redshift in the CPL, the ICPL, and the $\Lambda$CDM. The wave numbers $k$ in the three models are set as 1.0 (black line), 0.5 (dash green line), and 0.1 (dash red line), respectively.}
\end{figure}

\begin{figure}
	\centering
  \includegraphics[width=0.45\textwidth]{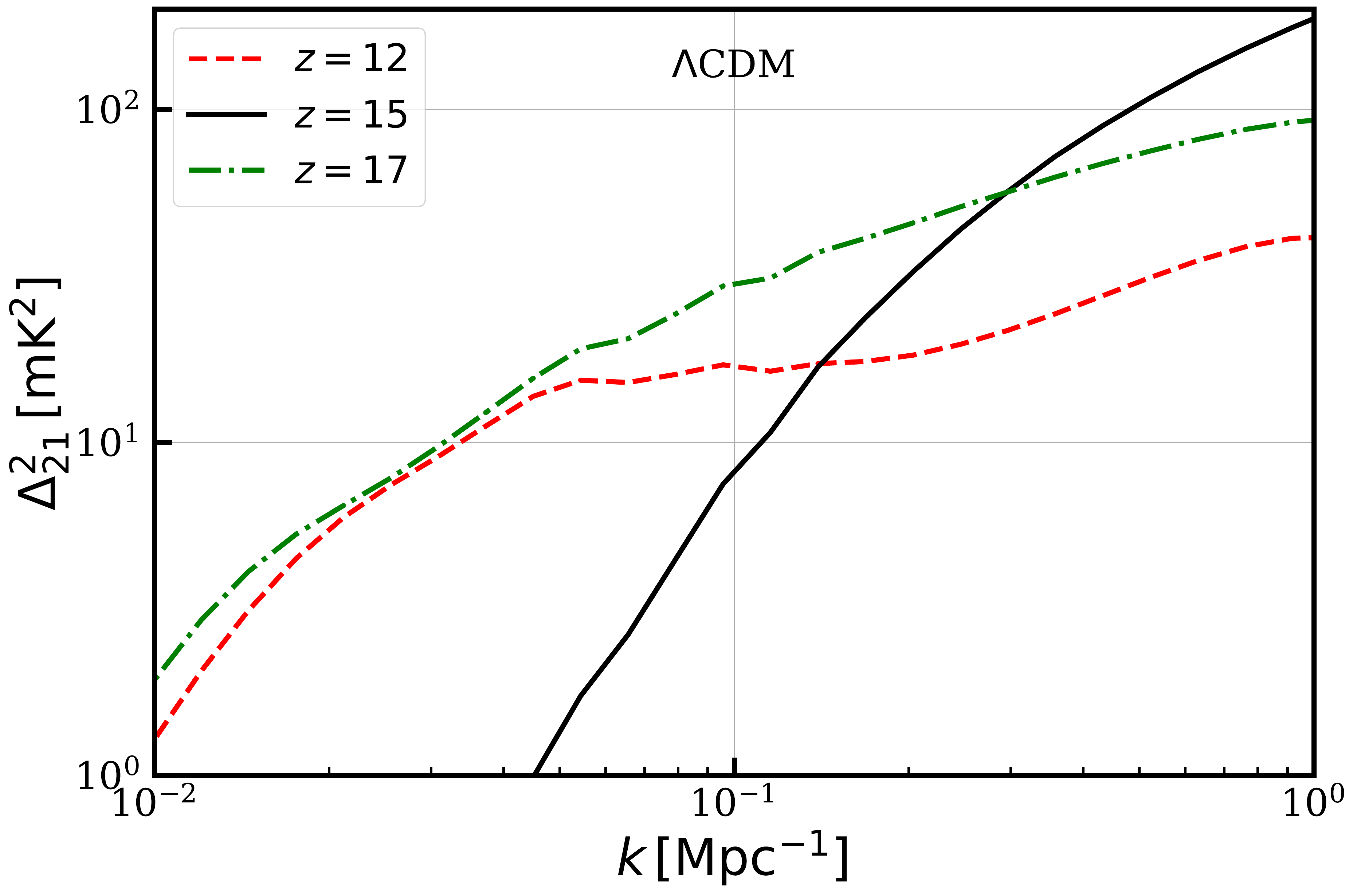}
  \includegraphics[width=0.45\textwidth]{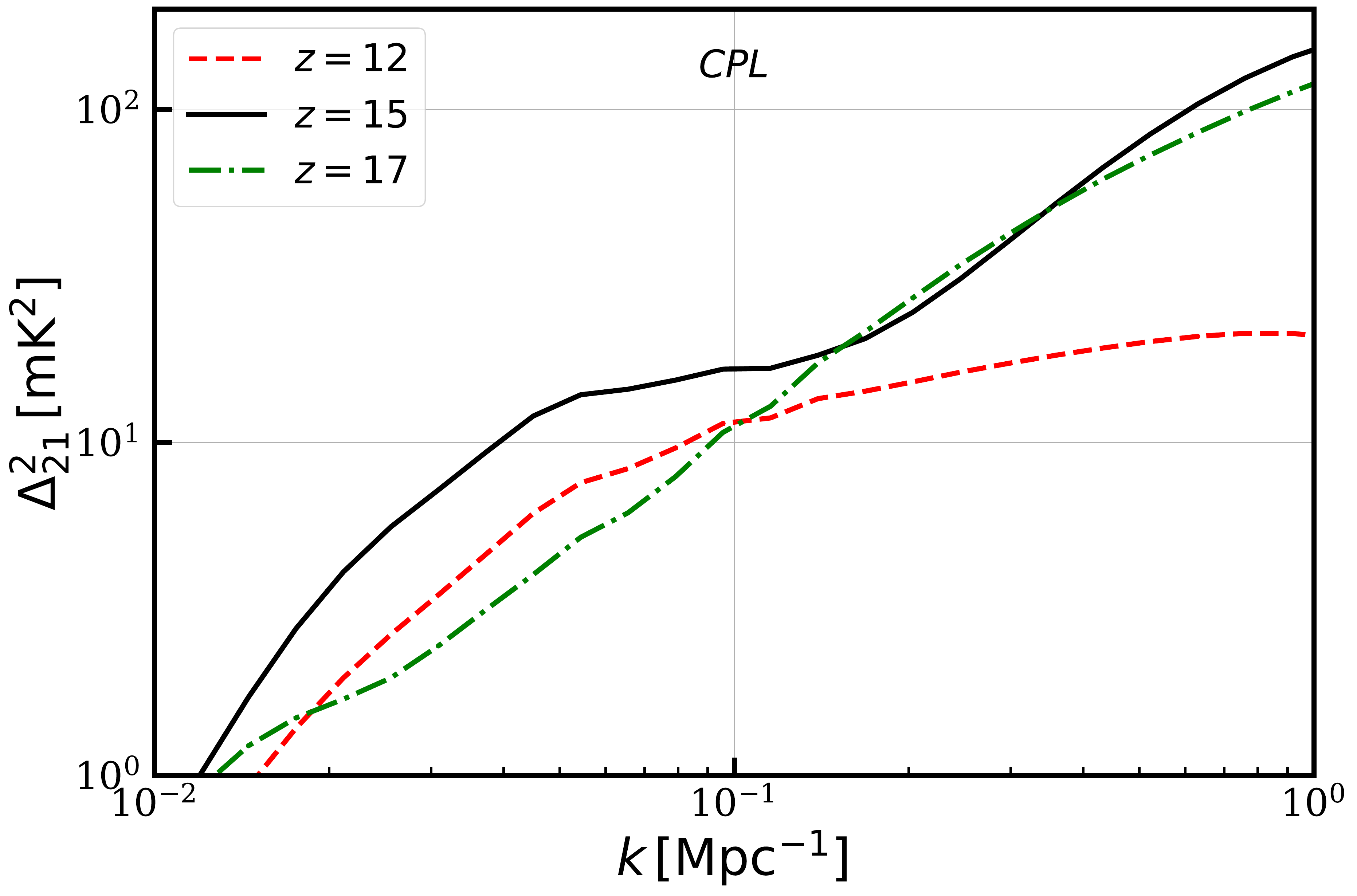}
  \includegraphics[width=0.45\textwidth]{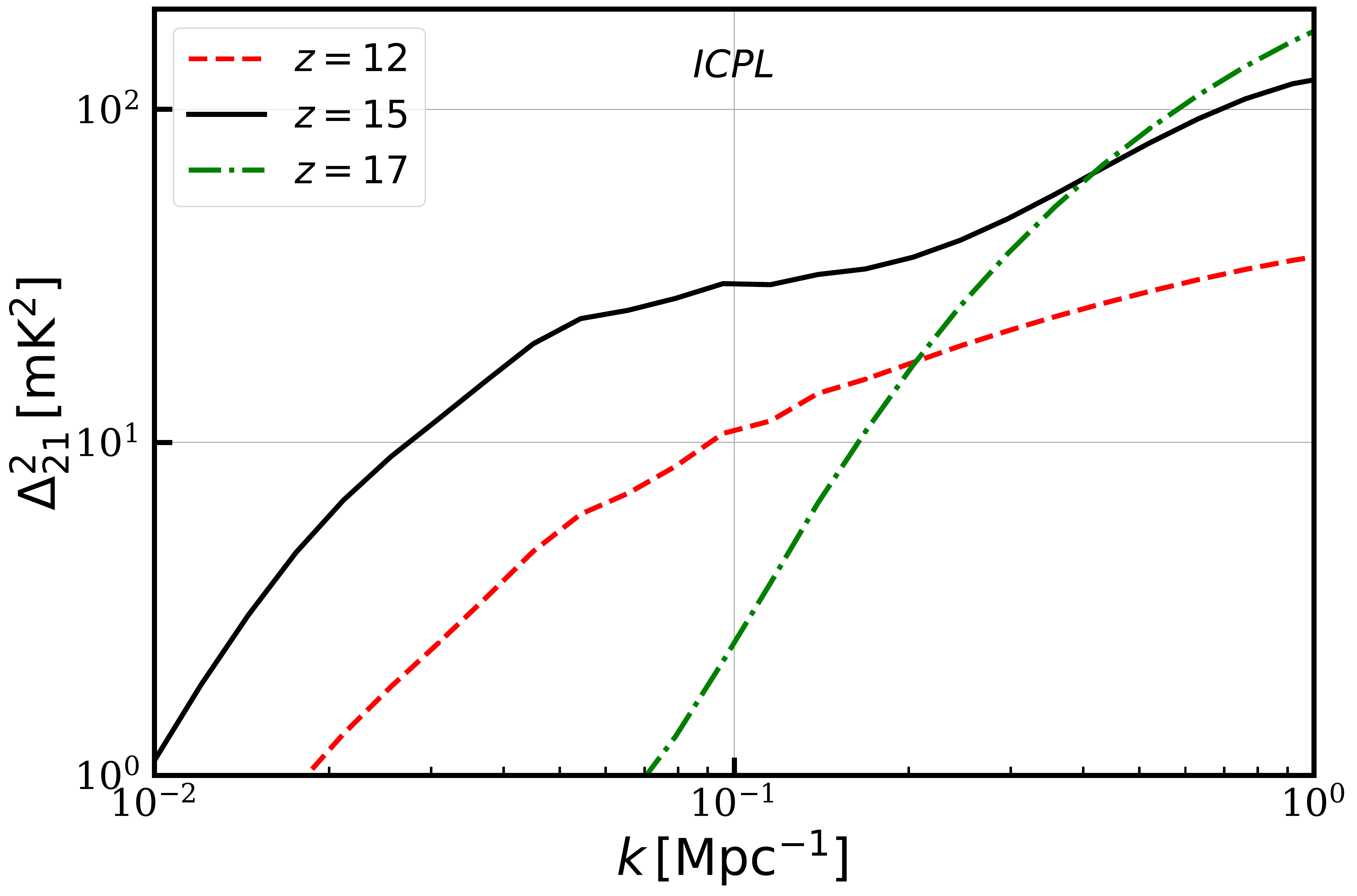}
	\caption{\label{fig:3}The evolution of 21-cm power spectra in the $\Lambda$CDM, the CPL, and hte ICPL model. The slices of redshift are selected at 12 (red dash line), 15 (black line), and 17 (green dash line), respectively.}
\end{figure}

\begin{figure}
	\centering
\includegraphics[width=0.7\textwidth]{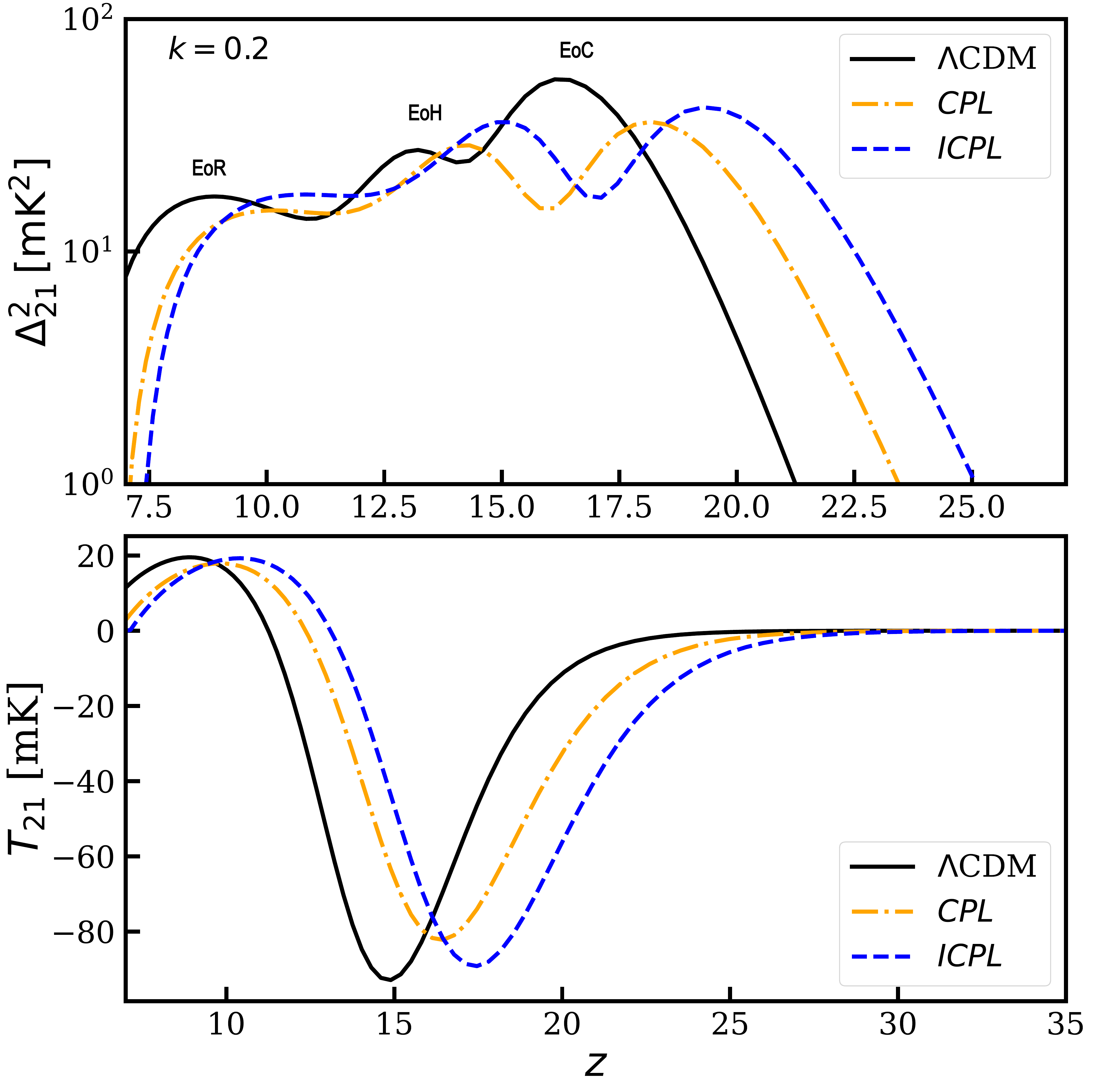}

	\caption{\label{fig:k02}The power spectra ($\Delta_{21}^2$) and brightness temperature $(T_{21})$ of the 21-cm signal. The theoretical value of the $\Lambda$CDM, the CPL, and the ICPL models are shown in the black line, yellow-dash line, and blue-dash line, respectively. In this figure, the wave number $k$ is 0.2.}
\end{figure}

\begin{figure}
	\centering
\includegraphics[width=0.7\textwidth]{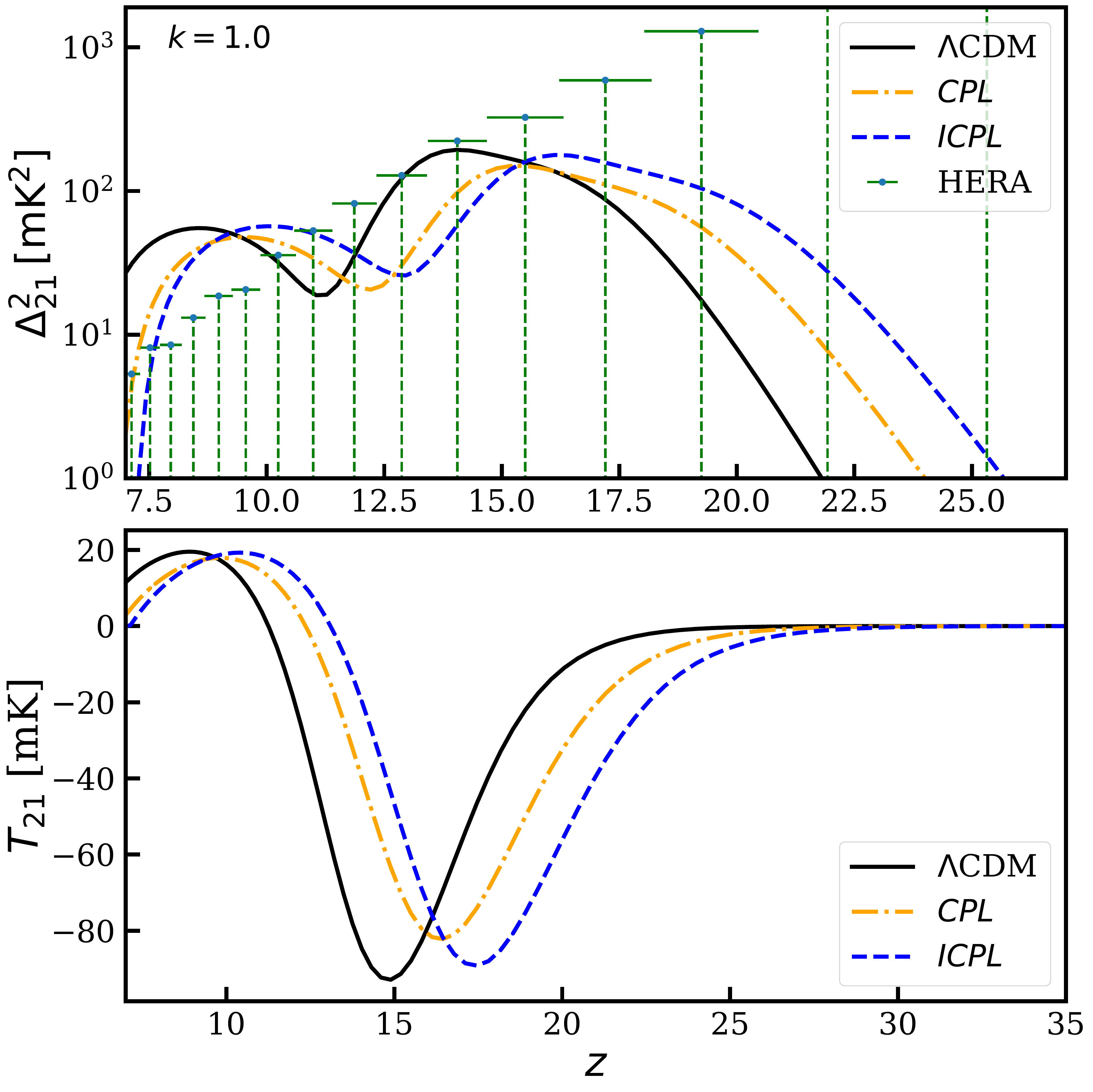}
	\caption{\label{fig:k1} The power spectra ($\Delta_{21}^2$) and brightness temperature $(T_{21})$ of the 21-cm signal. The theoretical value of the $\Lambda$CDM, the CPL, and the ICPL models are shown in the black line, yellow-dash line, and blue-dash line, respectively. In this figure, the wave number $k$ is 1.0. The expected noise from HERA shows in green points.}
\end{figure}

The three dark energy models above have been constrained by the previous research \cite{Chevallier:2000qy, Linder:2002et, Colgain:2021pmf, He:2008tn}. In this work, we are more interested in the evolution of the three models in the local Universe, so we fit free parameters by comparing the data from baryon acoustic oscillations (BAO) \cite{BOSS:2016wmc, BOSS:2016apd, Vargas-Magana:2016imr, BOSS:2016hvq} and Type Ia supernovae (SNIa) \cite{Pan-STARRS1:2017jku, Riess:2019cxk}. The best-fit results are shown in Tab.~\ref{tab:2}. 
$H_0$ value from the model and the ICPL model are $70.6986^{+0.2639}_{-0.2601} \mathrm{km/s/Mpc}$ and $70.9338^{+0.2549}_{-0.2515} \mathrm{km/s/Mpc}$. 
%Both models can release the $H_0$ tension better than that  in $\Lambda$CDM ($H_0= 69.9182^{+0.1398}_{-0.1369}\mathrm{km/s/Mpc}$).  
The fitting result of $\gamma$ is a negative number $-0.0078^{+0.0192}_{-0.0118}$, which means that a high possibility of the energy transfer from dark matter to dark energy.  
%In the following study, we shall use these constraint results from the 3 cases to calculate the global signal of 21-cm.

To get the 21-cm signal, we utilize the local SFRD  method to derive correlation functions incorporating nonlinear effects \cite{Munoz:2023kkg}. The SFRD serves as the fundamental component for determining the statistics of radiative fields by the $Zeus21$ program \cite{Munoz:2023kkg} combined with $CLASS$ \cite{Lesgourgues:2011re, Blas:2011rf}.
%, including X-ray and Lyman-alpha fluxes, and subsequently the 21-cm signal. To facilitate this analysis, we have developed the publicly available Python package called Zeus21. 
%This code enables us to accurately predict the 21-cm global signal and power spectrum within approximately several seconds, with minimal memory usage compared with other corresponding public programs. 
%SFRD can efficiently calculate any two-point function, for example, the power spectrum.
%--------fig 1
From Eq.~(\ref{eq:5}), we have known the 21-cm line signal is influenced by various factors, including the temperature of the cosmic microwave background $T_{\mathrm{CMB}}$, the kinetic temperature of the gas $T_k$, and the spin temperature $T_S$. From the evolution of these temperature components in Fig.~\ref{fig:1},
%, namely CMB temperature ($T_{cmb}$), gas kinetic temperature ($T_k$), and 21cm brightness temperature ($T_{21}$), can be observed in Figure \ref{fig:1}.
%Comparing the models of DDE and $\Lambda$CDM, 
we find that the spin temperature and gas kinetic temperature in the CPL and ICPL models are higher than the $\Lambda$CDM at redshift $z\approx 10$. 
%------------This can be attributed to the faster evolution of the two DDE models at cosmic dawn than $\Lambda$CDM (as indicated by the expansion rate (Hz) shown in the graph). 
$T_{\mathrm{CMB}}$ is nearly the same in all three models, which is easy to understand that dynamic dark energy models primarily affect the Universe locally while having less impact on the CMB temperature.
%This consistency aligns with the understanding that DDE models primarily affect the Universe after CMB while having less impact on the CMB temperature.

%-------fig 2
The power spectrum of the primordial density perturbations can be described as
\begin{equation}
    \Delta_{\mathrm{21}}^2=k^3\frac{P_{\mathrm{21}}(k)}{2\pi^2}.
\end{equation}
% $P(k)$ is the dimensionless power spectrum, which gives the amplitude of the density perturbations as a function of the wavenumber $k$. The quantity $\Delta^2$ represents the variance in the density fluctuations at a given scale $k$, and is proportional to the square of the amplitude of the Fourier transform of the perturbations.
$P_{\mathrm{21}}(k)$ is the 21-cm power spectrum, which measures the amplitude of the brightness temperature fluctuations as a function of spatial scale. 
%The power spectrum is a statistical tool commonly used in cosmology to describe the distribution of matter and radiation in the universe.
The quantity $\Delta_{21}^2(k)$ relates to the variance of the brightness temperature fluctuations on a given scale $k$ as 
$
    \langle\delta T_b(\mathbf{k})\delta T_b(\mathbf{k}')\rangle = (2\pi)^3\delta_\mathrm{D}(\mathbf{k}-\mathbf{k}')\Delta_{21}^2(k).
$
%$\langle\delta T_b(\mathbf{k})\delta T_b(\mathbf{k}')\rangle = (2\pi)^3\delta_\mathrm{D}(\mathbf{k}-\mathbf{k}')\Delta^2(k)$,
where $\delta T_{\mathrm{b}}(\mathbf{k})$ is the Fourier transform of the brightness temperature fluctuations, and $\delta_D(\mathbf{k})$ is the Dirac delta function. 
%The above equation shows that $\Delta^2(k)$ is a measure of the amount of power in the 21cm signal at a given spatial scale $k$.

We compare the $\Delta_{\mathrm{21}}^2$ of the three models in Fig.~\ref{fig:2}.  The scale factor $k$ in our setting is 0.1, 0.5, and 1.0, respectively. With these fixed values, we find that the power spectra increase with the value of $k$. 
%------delta_21
%In Figure 2, we present the evolution of the 21-cm power spectrum (PS) at 3 different Fourier-space wavenumbers, denoted as k and shown as 0.1, 0.5, and 1.0, respectively. The
At $k$ = 0.1 Mpc$^{-1}$, the power spectrum exhibits distinct features, characterized by two bumps from $z\geq 10$. The right bump corresponds to the epoch of coupling (EoC) which refers to the period when Lyman-alpha fluctuations dominate. During this epoch, the Lyman-alpha photons emitted by the first galaxies couple the spin temperature of neutral hydrogen to the kinetic temperature of the intergalactic medium (IGM) and this coupling is known as the WF effect. The left bump is the epoch of X-ray Heating (EoH). 
%It occurs when IGM temperature fluctuations dominate. 
It is characterized by the influence of X-rays emitted by galaxies, which heat up the surrounding IGM and affect the 21-cm signal. 
Between the two eras, the negative contribution from the cross-power between these fields leads to relative troughs in the power spectrum. 
%These features in the power spectrum provide valuable information about the underlying physical processes shaping the 21-cm signal during different cosmic epochs. 
Compared with the $\Lambda$CDM, the peak and concave point of the bump were shifted to an earlier Universe in CPL and ICPL. Especially, the ICPL has a higher corresponding redshift than CPL, which shows that the interaction term makes the ICPL obtain a faster evolution of our Universe.
% In between these eras, the negative contribution from the cross-power between these fields leads to relative troughs in the power spectrum. These features in the power spectrum provide valuable information about the underlying physical processes shaping the 21-cm signal during different cosmic epochs.
At smaller scales ($k$ = 1.0 Mpc$^{-1}$), the peak of the bump in the  $\Lambda$CDM is higher than the CPL and ICPL models. 
%the cancellation effect observed at larger scales does not occur, resulting in an overall increase in the power of the 21-cm signal. In comparison, the power is highest for the $\Lambda$CDM model at the peak of the bump. 
This is because the two models have smaller fluctuations in the 21-cm signal.

In Fig.~\ref{fig:3}, the  power spectrum  is shown at three specific redshifts. These 3 redshifts are chosen to represent different epoch slices in the $\Lambda$CDM: the EoH at $z=12$, the EoC at $z=17$, and the transition between EoH and EoC ($z=15$).
For $z=12$ and $z=17$, the power spectrum remains relatively flat with $k$. However, at the transition redshift of the $\Lambda$CDM, the  power spectra experience a significant drop. This drop  is caused by the negative contribution of cross-terms between the two physical processes.
%\red{Of particular interest is the redshift $z = 12$, where the power spectrum exhibits wiggles at a specific scale of $k$ = 0.04 Mpc$^{-1}$. These wiggles are attributed to the streaming velocities which possess acoustic oscillations. }
%These oscillations are imprinted onto the SFRD method through the feedback mechanisms, thereby affecting the 21-cm signal and introducing the observed wiggles in the power spectrum.
%The scale dependence of the 21-cm power spectrum provides valuable features into the different epochs and physical processes that shape the 21-cm signal.

Now, we plot the 21-cm power spectrum and the global brightness temperature of these three models with different wavenumber $k$ in Fig.~\ref{fig:k02} and Fig.~\ref{fig:k1}.
%, the wave number $k$ is fixed at $0.2$ and $1.0$, respectively. 
For $k=0.2$ in Fig.~\ref{fig:k02},  the third bump appears at $z$ around 9, which corresponds to the Epoch of Reionization (EoR). During this epoch, the first galaxies and quasars emit high-energy photons that ionize the neutral hydrogen in the intergalactic medium. As ionization spreads through the Universe, ionization fraction fluctuations become a dominant factor in shaping the 21-cm signal. The other bumps are the EoC and EoH, as discussed in Fig.~\ref{fig:2} and Fig.~\ref{fig:3}. Same with the previous results, the bump peaks in CPL and ICPL models were shifted to the earlier Universe, 
%It shows the ICPL model reaches this epoch of cosmic dawn earlier than the CPL model, and even earlier than $\Lambda$CDM. 
which indicates the formation of the first stars and galaxies occurs at ICPL earlier than the CPL model, and even earlier than the $\Lambda$CDM.
%These differences in the timing of cosmic dawn between different models highlight the impact of different cosmological parameters and assumptions on the evolution of the early universe and the formation of the first astrophysical objects.

The global 21-cm brightness temperature $T_{21}$ confirmed this result, %First, there is the EoC plotted as the flat line at high redshift, where the $T_{21}$ becomes more negative due to the WF coupling sourced by the Lyman-series photons from the first galaxies. 
%Then, there is the EoH from z around 17, where X-rays emitted by galaxies heat up the IGM, slowly increasing $T_{21}$ until it is above zero. Finally, during the EoR, $T_{21}$ is driven towards zero following the neutral hydrogen fraction ($x_{\mathrm{HI}}$). 
which the ICPL and CPL model enters every epoch earlier than the $\Lambda$CDM.
The lowest point of the absorption trough in the 21-cm signal, characterized by a temperature of approximately -90 mK, is significantly shallower compared to the value claimed by the EDGES experiment, which is around -500 mK \cite{Bowman:2018yin}. This significant discrepancy between the observations and the predicted signal has led to the exploration of alternative explanations.
%One possibility is to consider the presence of stronger electric charges for dark matter particles. This idea suggests that the interaction between dark matter and baryonic matter through electromagnetic forces could have a stronger impact on the 21-cm signal, leading to a shallower trough. Researchers such as Muñoz and Loeb (2018) and Barkana (2018) have proposed models incorporating stronger dark matter electric charges to explain the observed shallowness of the absorption trough.
%Another explanation involves considering a brighter additional radio background. The presence of an enhanced radio background would contribute additional radiation to the 21-cm signal, affecting its depth. Studies by Ewall-Wice et al. (2018) and Pospelov et al. (2018) have explored the possibility of a brighter radio background as an alternative explanation for the discrepancy between the observed and predicted 21-cm signals.
%These alternative explanations highlight the ongoing efforts to reconcile the observed 21-cm signal with theoretical predictions and 
The ICPL has the potential to enlarge the signal in the theory, but this phenomenon still needs to consider additional physical processes to affect the evolution of the intergalactic medium and the  21-cm signal.

We also considered the expected noise level of observations in Fig. \ref{fig:k1}.
It includes the expected noise level after 1 year (1080 hours) of integration with the HERA telescope shown in green points \cite{Munoz:2021psm}. The noise calculation is performed using the $\it{21cmSense}$ software \cite{Pober:2012zz, Pober:2013jna}, assuming a moderate-foreground scenario with a buffer value of $a = 0.1 h$Mpc$^{-1}$ above the horizon. 
%The buffer value is used to account for foreground contamination in the observations.
The noise level provides an estimate of the sensitivity and precision of the measurements that can be achieved with HERA under these assumptions. 
In our study, we demonstrate that the significance of the three models can be detected at redshift around $z \approx 10-15$. These features will help to distinguish the three models with more accurate observations in the future and will help us to distinguish the equation of state of dynamical dark energy.
%This provides a unique opportunity to probe the Universe and study the effects of relative velocities on the thermal history of the intergalactic medium.

%\begin{table}
%	\begin{center}
%	\caption{{\color{black}Best-fit of free parameters in $\Lambda$CDM, $w$CDM, I$w$CDM, respectively. These results with 68$\%$ C.L. based on the observation of CMB, BAO, and Type Ia Supernova.}}
%	\begin{tabular}{|l|c|c|}
%		 \hline $\text { Parameter } $& $w$CDM & I$w$CDM \\
%		\hline 
%            $H_0 $      & $70.71^{+0.26}_{-0.27} $      &$ 70.73\pm{0.25}$\\
%            \hline
%		$\Omega_b $ & $ 0.0433\pm 0.0005 $          &$0.0433 \pm 0.0005 $\\
%  \hline
%		$\Omega_{c}$& $ 0.2723^{+0.0103}_{-0.0100}$ &$ 0.2698^{+0.0100}_{-0.0102} $\\
%  \hline
%		$w_0 $        & $-1.217^{+0.069}_{-0.067} $   &$-1.226^{+0.068}_{-0.064}$\\
%  \hline
%		$w_a$        & $0.675^{+0.380}_{-0.405}$     &$ 1.420^{+0.516}_{-0.968}$ \\
%  \hline
%		$\gamma$    &$ 0 $&$0.150^{+0.334}_{-0.170}$\\
%		\hline
% 		\end{tabular}
%\label{tab:old}
%\end{center}
%\end{table}

\section{Summary}
\label{sec:5}

In this paper, we discussed the difference in the background evolution in the $\Lambda$CDM, the CPL model, and the ICPL model. The ICPL model was considered with an interaction term  of the form $Q=3\gamma H\rho_c$.
%, where $\gamma$ is a coupling parameter and $\rho_c$ is the  energy density of dark matter. 
%We calculated the evolution of this model and obtained the best-fit parameters through comparison with observational data from Type Ia Supernovae (SN) and Baryon Acoustic Oscillations (BAO).
We numerically compared the best-fit parameters obtained for the three models with Type Ia Supernovae and Baryon Acoustic Oscillations data, which are the local observations without being disturbed by the CMB. The result of $\gamma$ is a negative number, that shows the energy transfer from dark matter to dark energy.
%The results are presented in Table \ref{tab:2}. It is worth noting that in our analysis, we did not consider the Cosmic Microwave Background observations. Instead, we focused solely on the local Universe simulation to study the 21-cm global signal.
These results were used in the calculation of 21-cm global signal.

We calculated the evolution of the CMB temperature $T_{\mathrm{CMB}}$, kinetic temperature $T_k$, and spin temperature $T_s$ in different cosmological models with the $Zeus21$ and $\textit{CLASS}$ program. We found that $T_{\mathrm{CMB}}$ did not exhibit significant changes among the three models. However, the $T_k$ and $T_S$ values in the CPL and ICPL models were approximately twice as large as those in the $\Lambda$CDM.
Moreover, we analyzed the 21-cm power spectra at various wave numbers $k$, to investigate the epoch of reionization, the epoch of X-ray heating, and the epoch of coupling in the three models.
We found that these epochs in the CPL and ICPL models occurred at earlier redshifts due to the different expansion rates during cosmic dawn. 
The variations of bumps in the power spectra provided valuable information about the process of reionization.

We also plotted the brightness temperature of the 21-cm line of the three models. We found that the moment of the first galaxy formation and the epoch of reheating occurred earlier in CPL and ICPL models compared to the $\Lambda$CDM. Specifically, the ICPL model exhibited an even earlier onset than the CPL model.
In our calculation, the ICPL model can help to relieve the tension about the lowest point of the absorption between the theoretical expectations and the strength of this signal reported by the EDGES experiment, which shows that the interacting dark energy model has the potential to get a lower $T_{21}$ signal.

Furthermore, based on our analysis using HERA expected noise signal, we find that the features of the $\Lambda$CDM, the CPL, and the ICPL model have possibly to be detected 
%and distinguished at redshifts around $z \approx 10-15$ 
with more accurate observation.  This suggests that future observations with HERA have the potential to distinguish different dark energy models and their equations of state. The observation of 21-cm can provide valuable information on the nature of dark energy and the early Universe.

\section*{Acknowledgements}

We thank Julian B. Mu\~{n}oz 
%{, Jes\'us Cruz Rojas}
for the useful discussion. 
LY was supported by the YST project in APCTP.


\begin{thebibliography}{100}
	
%	\scriptsize
%	\footnotesize
	

%\cite{Bowman:2018yin}
\bibitem{Bowman:2018yin}
J.~D.~Bowman, A.~E.~E.~Rogers, R.~A.~Monsalve, T.~J.~Mozdzen and N.~Mahesh,
%``An absorption profile centred at 78 megahertz in the sky-averaged spectrum,''
Nature \textbf{555}, no.7694, 67-70 (2018)
doi:10.1038/nature25792
[arXiv:1810.05912 [astro-ph.CO]].
%749 citations counted in INSPIRE as of 29 May 2023

%\cite{Pritchard:2011xb}
\bibitem{Pritchard:2011xb}
J.~R.~Pritchard and A.~Loeb,
%``21-cm cosmology,''
Rept. Prog. Phys. \textbf{75}, 086901 (2012)
doi:10.1088/0034-4885/75/8/086901
[arXiv:1109.6012 [astro-ph.CO]].
%588 citations counted in INSPIRE as of 12 May 2023

%\cite{Morales:2009gs}
\bibitem{Morales:2009gs}
M.~F.~Morales and J.~S.~B.~Wyithe,
%``Reionization and Cosmology with 21 cm Fluctuations,''
Ann. Rev. Astron. Astrophys. \textbf{48}, 127-171 (2010)
doi:10.1146/annurev-astro-081309-130936
[arXiv:0910.3010 [astro-ph.CO]].
%321 citations counted in INSPIRE as of 25 Apr 2023

%\cite{Furlanetto:2006jb}
\bibitem{Furlanetto:2006jb}
S.~Furlanetto, S.~P.~Oh and F.~Briggs,
%``Cosmology at Low Frequencies: The 21 cm Transition and the High-Redshift Universe,''
Phys. Rept. \textbf{433}, 181-301 (2006)
doi:10.1016/j.physrep.2006.08.002
[arXiv:astro-ph/0608032 [astro-ph]].
%1000 citations counted in INSPIRE as of 26 May 2023

%\cite{Tashiro:2014tsa}
\bibitem{Tashiro:2014tsa}
H.~Tashiro, K.~Kadota and J.~Silk,
%``Effects of dark matter-baryon scattering on redshifted 21 cm signals,''
Phys. Rev. D \textbf{90}, no.8, 083522 (2014)
doi:10.1103/PhysRevD.90.083522
[arXiv:1408.2571 [astro-ph.CO]].
%104 citations counted in INSPIRE as of 16 May 2023

%\cite{Feng:2018rje}
\bibitem{Feng:2018rje}
C.~Feng and G.~Holder,
%``Enhanced global signal of neutral hydrogen due to excess radiation at cosmic dawn,''
Astrophys. J. Lett. \textbf{858}, no.2, L17 (2018)
doi:10.3847/2041-8213/aac0fe
[arXiv:1802.07432 [astro-ph.CO]].
%197 citations counted in INSPIRE as of 29 May 2023

%\cite{Barkana:2018qrx}
\bibitem{Barkana:2018qrx}
R.~Barkana, N.~J.~Outmezguine, D.~Redigolo and T.~Volansky,
%``Strong constraints on light dark matter interpretation of the EDGES signal,''
Phys. Rev. D \textbf{98}, no.10, 103005 (2018)
doi:10.1103/PhysRevD.98.103005
[arXiv:1803.03091 [hep-ph]].
%208 citations counted in INSPIRE as of 26 May 2023

%\cite{Mahdawi:2018euy}
\bibitem{Mahdawi:2018euy}
M.~S.~Mahdawi and G.~R.~Farrar,
%``Constraints on Dark Matter with a moderately large and velocity-dependent DM-nucleon cross-section,''
JCAP \textbf{10}, 007 (2018)
doi:10.1088/1475-7516/2018/10/007
[arXiv:1804.03073 [hep-ph]].
%69 citations counted in INSPIRE as of 15 Mar 2023

%\cite{Mirocha:2018cih}
\bibitem{Mirocha:2018cih}
J.~Mirocha and S.~R.~Furlanetto,
%``What does the first highly-redshifted 21-cm detection tell us about early galaxies?,''
Mon. Not. Roy. Astron. Soc. \textbf{483}, no.2, 1980-1992 (2019)
doi:10.1093/mnras/sty3260
[arXiv:1803.03272 [astro-ph.GA]].
%147 citations counted in INSPIRE as of 26 May 2023

%\cite{Ewall-Wice:2018bzf}
\bibitem{Ewall-Wice:2018bzf}
A.~Ewall-Wice, T.~C.~Chang, J.~Lazio, O.~Dore, M.~Seiffert and R.~A.~Monsalve,
%``Modeling the Radio Background from the First Black Holes at Cosmic Dawn: Implications for the 21 cm Absorption Amplitude,''
Astrophys. J. \textbf{868}, no.1, 63 (2018)
doi:10.3847/1538-4357/aae51d
[arXiv:1803.01815 [astro-ph.CO]].
%183 citations counted in INSPIRE as of 29 May 2023

%\cite{Hirano:2018alc}
\bibitem{Hirano:2018alc}
S.~Hirano and V.~Bromm,
%``Baryon-dark matter scattering and first star formation,''
Mon. Not. Roy. Astron. Soc. \textbf{480}, no.1, L85-L89 (2018)
doi:10.1093/mnrasl/sly132
[arXiv:1803.10671 [astro-ph.GA]].
%27 citations counted in INSPIRE as of 15 Mar 2023

%\cite{Venumadhav:2018uwn}
\bibitem{Venumadhav:2018uwn}
T.~Venumadhav, L.~Dai, A.~Kaurov and M.~Zaldarriaga,
%``Heating of the intergalactic medium by the cosmic microwave background during cosmic dawn,''
Phys. Rev. D \textbf{98}, no.10, 103513 (2018)
doi:10.1103/PhysRevD.98.103513
[arXiv:1804.02406 [astro-ph.CO]].
%51 citations counted in INSPIRE as of 17 Apr 2023

%\cite{Clark:2018ghm}
\bibitem{Clark:2018ghm}
S.~Clark, B.~Dutta, Y.~Gao, Y.~Z.~Ma and L.~E.~Strigari,
%``21 cm limits on decaying dark matter and primordial black holes,''
Phys. Rev. D \textbf{98}, no.4, 043006 (2018)
doi:10.1103/PhysRevD.98.043006
[arXiv:1803.09390 [astro-ph.HE]].
%127 citations counted in INSPIRE as of 03 May 2023

%\cite{Munoz:2018jwq}
\bibitem{Munoz:2018jwq}
J.~B.~Mu\~noz, C.~Dvorkin and A.~Loeb,
%``21-cm Fluctuations from Charged Dark Matter,''
Phys. Rev. Lett. \textbf{121}, no.12, 121301 (2018)
doi:10.1103/PhysRevLett.121.121301
[arXiv:1804.01092 [astro-ph.CO]].
%88 citations counted in INSPIRE as of 05 May 2023

%\cite{Pospelov:2018kdh}
\bibitem{Pospelov:2018kdh}
M.~Pospelov, J.~Pradler, J.~T.~Ruderman and A.~Urbano,
%``Room for New Physics in the Rayleigh-Jeans Tail of the Cosmic Microwave Background,''
Phys. Rev. Lett. \textbf{121}, no.3, 031103 (2018)
doi:10.1103/PhysRevLett.121.031103
[arXiv:1803.07048 [hep-ph]].
%125 citations counted in INSPIRE as of 16 May 2023

%\cite{Li:2018kzs}
\bibitem{Li:2018kzs}
C.~Li and Y.~F.~Cai,
%``Searching for the Dark Force with 21-cm Spectrum in Light of EDGES,''
Phys. Lett. B \textbf{788}, 70-75 (2019)
doi:10.1016/j.physletb.2018.11.011
[arXiv:1804.04816 [astro-ph.CO]].
%25 citations counted in INSPIRE as of 15 Mar 2023

%\cite{Hektor:2018lec}
\bibitem{Hektor:2018lec}
A.~Hektor, G.~H\"utsi, L.~Marzola and V.~Vaskonen,
%``Constraints on ALPs and excited dark matter from the EDGES 21-cm absorption signal,''
Phys. Lett. B \textbf{785}, 429-433 (2018)
doi:10.1016/j.physletb.2018.09.009
[arXiv:1805.09319 [hep-ph]].
%23 citations counted in INSPIRE as of 09 May 2023

%\cite{Li:2018okf}
\bibitem{Li:2018okf}
C.~Li, N.~Houston, T.~Li, Q.~Yang and X.~Zhang,
%``A detailed exploration of the EDGES 21cm absorption anomaly and axion-induced cooling,''
Int. J. Mod. Phys. D \textbf{30}, no.06, 2150041 (2021)
doi:10.1142/S0218271821500413
[arXiv:1812.03931 [hep-ph]].
%11 citations counted in INSPIRE as of 23 May 2023

%\cite{Fraser:2018acy}
\bibitem{Fraser:2018acy}
S.~Fraser, A.~Hektor, G.~H\"utsi, K.~Kannike, C.~Marzo, L.~Marzola, C.~Spethmann, A.~Racioppi, M.~Raidal and V.~Vaskonen, \textit{et al.}
%``The EDGES 21 cm Anomaly and Properties of Dark Matter,''
Phys. Lett. B \textbf{785}, 159-164 (2018)
doi:10.1016/j.physletb.2018.08.035
[arXiv:1803.03245 [hep-ph]].
%156 citations counted in INSPIRE as of 09 May 2023

%\cite{Yang:2018gjd}
\bibitem{Yang:2018gjd}
Y.~Yang,
%``Contributions of dark matter annihilation to the global 21 cm spectrum observed by the EDGES experiment,''
Phys. Rev. D \textbf{98}, no.10, 103503 (2018)
doi:10.1103/PhysRevD.98.103503
[arXiv:1803.05803 [astro-ph.CO]].
%33 citations counted in INSPIRE as of 03 May 2023

%\cite{DAmico:2018sxd}
\bibitem{DAmico:2018sxd}
G.~D'Amico, P.~Panci and A.~Strumia,
%``Bounds on Dark Matter annihilations from 21 cm data,''
Phys. Rev. Lett. \textbf{121}, no.1, 011103 (2018)
doi:10.1103/PhysRevLett.121.011103
[arXiv:1803.03629 [astro-ph.CO]].
%92 citations counted in INSPIRE as of 26 May 2023

%\cite{Mitridate:2018iag}
\bibitem{Mitridate:2018iag}
A.~Mitridate and A.~Podo,
%``Bounds on Dark Matter decay from 21 cm line,''
JCAP \textbf{05}, 069 (2018)
doi:10.1088/1475-7516/2018/05/069
[arXiv:1803.11169 [hep-ph]].
%60 citations counted in INSPIRE as of 15 Mar 2023

%\cite{Cheung:2018vww}
\bibitem{Cheung:2018vww}
K.~Cheung, J.~L.~Kuo, K.~W.~Ng and Y.~L.~S.~Tsai,
%``The impact of EDGES 21-cm data on dark matter interactions,''
Phys. Lett. B \textbf{789}, 137-144 (2019)
doi:10.1016/j.physletb.2018.11.058
[arXiv:1803.09398 [astro-ph.CO]].
%53 citations counted in INSPIRE as of 03 May 2023

%\cite{Barkana:2018lgd}
\bibitem{Barkana:2018lgd}
R.~Barkana,
%``Possible interaction between baryons and dark-matter particles revealed by the first stars,''
Nature \textbf{555}, no.7694, 71-74 (2018)
doi:10.1038/nature25791
[arXiv:1803.06698 [astro-ph.CO]].
%391 citations counted in INSPIRE as of 26 May 2023

%\cite{Kovetz:2018zan}
\bibitem{Kovetz:2018zan}
E.~D.~Kovetz, V.~Poulin, V.~Gluscevic, K.~K.~Boddy, R.~Barkana and M.~Kamionkowski,
%``Tighter limits on dark matter explanations of the anomalous EDGES 21 cm signal,''
Phys. Rev. D \textbf{98}, no.10, 103529 (2018)
doi:10.1103/PhysRevD.98.103529
[arXiv:1807.11482 [astro-ph.CO]].
%132 citations counted in INSPIRE as of 29 May 2023

%\cite{Slatyer:2018aqg}
\bibitem{Slatyer:2018aqg}
T.~R.~Slatyer and C.~L.~Wu,
%``Early-Universe constraints on dark matter-baryon scattering and their implications for a global 21 cm signal,''
Phys. Rev. D \textbf{98}, no.2, 023013 (2018)
doi:10.1103/PhysRevD.98.023013
[arXiv:1803.09734 [astro-ph.CO]].
%169 citations counted in INSPIRE as of 16 May 2023

%\cite{Munoz:2018pzp}
\bibitem{Munoz:2018pzp}
J.~B.~Mu\~noz and A.~Loeb,
%``A small amount of mini-charged dark matter could cool the baryons in the early Universe,''
Nature \textbf{557}, no.7707, 684 (2018)
doi:10.1038/s41586-018-0151-x
[arXiv:1802.10094 [astro-ph.CO]].
%251 citations counted in INSPIRE as of 23 May 2023

%\cite{Berlin:2018sjs}
\bibitem{Berlin:2018sjs}
A.~Berlin, D.~Hooper, G.~Krnjaic and S.~D.~McDermott,
%``Severely Constraining Dark Matter Interpretations of the 21-cm Anomaly,''
Phys. Rev. Lett. \textbf{121}, no.1, 011102 (2018)
doi:10.1103/PhysRevLett.121.011102
[arXiv:1803.02804 [hep-ph]].
%224 citations counted in INSPIRE as of 11 May 2023

%\cite{Xiao:2018jyl}
\bibitem{Xiao:2018jyl}
L.~Xiao, R.~An, L.~Zhang, B.~Yue, Y.~Xu and B.~Wang,
%``Can conformal and disformal couplings between dark sectors explain the EDGES 21-cm anomaly?,''
Phys. Rev. D \textbf{99}, no.2, 023528 (2019)
doi:10.1103/PhysRevD.99.023528
[arXiv:1807.05541 [astro-ph.CO]].
%22 citations counted in INSPIRE as of 15 Mar 2023

%\cite{Costa:2018aoy}
\bibitem{Costa:2018aoy}
A.~A.~Costa, R.~C.~G.~Landim, B.~Wang and E.~Abdalla,
%``Interacting Dark Energy: Possible Explanation for 21-cm Absorption at Cosmic Dawn,''
Eur. Phys. J. C \textbf{78}, no.9, 746 (2018)
doi:10.1140/epjc/s10052-018-6237-7
[arXiv:1803.06944 [astro-ph.CO]].
%70 citations counted in INSPIRE as of 15 Mar 2023

%\cite{Wang:2018azy}
\bibitem{Wang:2018azy}
Y.~Wang and G.~B.~Zhao,
%``Constraining the dark matter-vacuum energy interaction using the EDGES 21-cm absorption signal,''
Astrophys. J. \textbf{869}, no.1, 26 (2018)
doi:10.3847/1538-4357/aaeb9c
[arXiv:1805.11210 [astro-ph.CO]].
%24 citations counted in INSPIRE as of 31 Mar 2023

%\cite{Li:2019loh}
\bibitem{Li:2019loh}
C.~Li, X.~Ren, M.~Khurshudyan and Y.~F.~Cai,
%``Implications of the possible 21-cm line excess at cosmic dawn on dynamics of interacting dark energy,''
Phys. Lett. B \textbf{801}, 135141 (2020)
doi:10.1016/j.physletb.2019.135141
[arXiv:1904.02458 [astro-ph.CO]].
%41 citations counted in INSPIRE as of 15 Mar 2023

%\cite{Chevallier:2000qy}
\bibitem{Chevallier:2000qy}
M.~Chevallier and D.~Polarski,
%``Accelerating universes with scaling dark matter,''
Int. J. Mod. Phys. D \textbf{10}, 213-224 (2001)
doi:10.1142/S0218271801000822
[arXiv:gr-qc/0009008 [gr-qc]].
%1885 citations counted in INSPIRE as of 31 May 2023

%\cite{Linder:2002et}
\bibitem{Linder:2002et}
E.~V.~Linder,
%``Exploring the expansion history of the universe,''
Phys. Rev. Lett. \textbf{90}, 091301 (2003)
doi:10.1103/PhysRevLett.90.091301
[arXiv:astro-ph/0208512 [astro-ph]].
%1783 citations counted in INSPIRE as of 31 May 2023

%\cite{Colgain:2021pmf}
\bibitem{Colgain:2021pmf}
E.~\'O.~Colg\'ain, M.~M.~Sheikh-Jabbari and L.~Yin,
%``Can dark energy be dynamical?,''
Phys. Rev. D \textbf{104}, no.2, 023510 (2021)
doi:10.1103/PhysRevD.104.023510
[arXiv:2104.01930 [astro-ph.CO]].
%29 citations counted in INSPIRE as of 18 Apr 2023

%\cite{He:2008tn}
\bibitem{He:2008tn}
J.~H.~He and B.~Wang,
%``Effects of the interaction between dark energy and dark matter on cosmological parameters,''
JCAP \textbf{06}, 010 (2008)
doi:10.1088/1475-7516/2008/06/010
[arXiv:0801.4233 [astro-ph]].
%219 citations counted in INSPIRE as of 19 Apr 2023

%\cite{Munoz:2023kkg}
\bibitem{Munoz:2023kkg}
J.~B.~Mu\~noz,
%``An Effective Model for the Cosmic-Dawn 21-cm Signal,''
[arXiv:2302.08506 [astro-ph.CO]].
%2 citations counted in INSPIRE as of 16 May 2023

%\cite{Singh:2017syr}
\bibitem{Singh:2017syr}
S.~Singh, R.~Subrahmanyan, N.~U.~Shankar, M.~S.~Rao, B.~S.~Girish, A.~Raghunathan, R.~Somashekar and K.~S.~Srivani,
%``SARAS 2: A Spectral Radiometer for probing Cosmic Dawn and the Epoch of Reionization through detection of the global 21 cm signal,''
Exper. Astron. \textbf{45}, no.2, 269-314 (2018)
doi:10.1007/s10686-018-9584-3
[arXiv:1710.01101 [astro-ph.IM]].
%52 citations counted in INSPIRE as of 04 May 2023

%\cite{Peitzmann:1996xv}
\bibitem{Peitzmann:1996xv}
T.~Peitzmann, C.~Barlag, C.~Blume, E.~M.~Bohne, D.~Bucher, A.~Claussen, R.~Glasow, N.~Heine, K.~H.~Kampert and R.~Santo, \textit{et al.}
%``A New monitoring system for the photon spectrometer LEDA in the WA98 experiment,''
Nucl. Instrum. Meth. A \textbf{376}, 368-374 (1996)
doi:10.1016/0168-9002(96)00224-0
%17 citations counted in INSPIRE as of 15 Mar 2023

%\cite{vanHaarlem:2013dsa}
\bibitem{vanHaarlem:2013dsa}
M.~P.~van Haarlem, M.~W.~Wise, A.~W.~Gunst, G.~Heald, J.~P.~McKean, J.~W.~T.~Hessels, A.~G.~de Bruyn, R.~Nijboer, J.~Swinbank and R.~Fallows, \textit{et al.}
%``LOFAR: The LOw-Frequency ARray,''
Astron. Astrophys. \textbf{556}, A2 (2013)
doi:10.1051/0004-6361/201220873
[arXiv:1305.3550 [astro-ph.IM]].
%868 citations counted in INSPIRE as of 24 May 2023

%\cite{Beardsley:2016njr}
\bibitem{Beardsley:2016njr}
A.~P.~Beardsley, B.~J.~Hazelton, I.~S.~Sullivan, P.~Carroll, N.~Barry, M.~Rahimi, B.~Pindor, C.~M.~Trott, J.~Line and D.~C.~Jacobs, \textit{et al.}
%``First Season MWA EoR Power Spectrum Results at Redshift 7,''
Astrophys. J. \textbf{833}, no.1, 102 (2016)
doi:10.3847/1538-4357/833/1/102
[arXiv:1608.06281 [astro-ph.IM]].
%145 citations counted in INSPIRE as of 30 Mar 2023

%\cite{Tingay:2012ps}
\bibitem{Tingay:2012ps}
S.~J.~Tingay, R.~Goeke, J.~D.~Bowman, D.~Emrich, S.~M.~Ord, D.~A.~Mitchell, M.~F.~Morales, T.~Booler, B.~Crosse and D.~Pallot, \textit{et al.}
%``The Murchison Widefield Array: the Square Kilometre Array Precursor at low radio frequencies,''
Publ. Astron. Soc. Austral. \textbf{30}, 7 (2013)
doi:10.1017/pasa.2012.007
[arXiv:1206.6945 [astro-ph.IM]].
%429 citations counted in INSPIRE as of 28 Apr 2023

%\cite{DeBoer:2016tnn}
\bibitem{DeBoer:2016tnn}
D.~R.~DeBoer, A.~R.~Parsons, J.~E.~Aguirre, P.~Alexander, Z.~S.~Ali, A.~P.~Beardsley, G.~Bernardi, J.~D.~Bowman, R.~F.~Bradley and C.~L.~Carilli, \textit{et al.}
%``Hydrogen Epoch of Reionization Array (HERA),''
Publ. Astron. Soc. Pac. \textbf{129}, no.974, 045001 (2017)
doi:10.1088/1538-3873/129/974/045001
[arXiv:1606.07473 [astro-ph.IM]].
%422 citations counted in INSPIRE as of 15 May 2023

%\cite{HERA:2021noe}
\bibitem{HERA:2021noe}
Z.~Abdurashidova \textit{et al.} [HERA],
%``HERA Phase I Limits on the Cosmic 21 cm Signal: Constraints on Astrophysics and Cosmology during the Epoch of Reionization,''
Astrophys. J. \textbf{924}, no.2, 51 (2022)
doi:10.3847/1538-4357/ac2ffc
[arXiv:2108.07282 [astro-ph.CO]].
%68 citations counted in INSPIRE as of 26 May 2023

%\cite{Santos:2015gra}
\bibitem{Santos:2015gra}
M.~G.~Santos, P.~Bull, D.~Alonso, S.~Camera, P.~G.~Ferreira, G.~Bernardi, R.~Maartens, M.~Viel, F.~Villaescusa-Navarro and F.~B.~Abdalla, \textit{et al.}
%``Cosmology from a SKA HI intensity mapping survey,''
PoS \textbf{AASKA14}, 019 (2015)
doi:10.22323/1.215.0019
[arXiv:1501.03989 [astro-ph.CO]].
%126 citations counted in INSPIRE as of 25 May 2023

%\cite{Zhang:2019dyq}
\bibitem{Zhang:2019dyq}
J.~F.~Zhang, L.~Y.~Gao, D.~Z.~He and X.~Zhang,
%``Improving cosmological parameter estimation with the future 21 cm observation from SKA,''
Phys. Lett. B \textbf{799} (2019), 135064
doi:10.1016/j.physletb.2019.135064
[arXiv:1908.03732 [astro-ph.CO]].
%13 citations counted in INSPIRE as of 31 May 2023

%\cite{Xu:2020uws}
\bibitem{Xu:2020uws}
Y.~Xu and X.~Zhang,
%``Cosmological parameter measurement and neutral hydrogen 21 cm sky survey with the Square Kilometre Array,''
Sci. China Phys. Mech. Astron. \textbf{63} (2020) no.7, 270431
doi:10.1007/s11433-020-1544-3
[arXiv:2002.00572 [astro-ph.CO]].
%18 citations counted in INSPIRE as of 31 May 2023

%\cite{Wang:2022oou}
\bibitem{Wang:2022oou}
L.~F.~Wang, Y.~Shao, G.~P.~Zhang, J.~F.~Zhang and X.~Zhang,
%``Ultra-low-frequency gravitational waves from individual supermassive black hole binaries as standard sirens,''
[arXiv:2201.00607 [astro-ph.CO]].
%7 citations counted in INSPIRE as of 31 May 2023

%\cite{Hartley:2023ach}
\bibitem{Hartley:2023ach}
P.~Hartley, A.~Bonaldi, R.~Braun, J.~N.~H.~S.~Aditya, S.~Aicardi, L.~Alegre, A.~Chakraborty, X.~Chen, S.~Choudhuri and A.~O.~Clarke, \textit{et al.}
%``SKA Science Data Challenge 2: analysis and results,''
doi:10.1093/mnras/stad1375
[arXiv:2303.07943 [astro-ph.IM]].
%0 citations counted in INSPIRE as of 31 May 2023
%In this work, we consider the phenomenology of these 

%\cite{Barkana:2000fd}
\bibitem{Barkana:2000fd}
R.~Barkana and A.~Loeb,
%``In the beginning: The First sources of light and the reionization of the Universe,''
Phys. Rept. \textbf{349}, 125-238 (2001)
doi:10.1016/S0370-1573(01)00019-9
[arXiv:astro-ph/0010468 [astro-ph]].
%944 citations counted in INSPIRE as of 18 May 2023

%\cite{Barkana:2004vb}
\bibitem{Barkana:2004vb}
R.~Barkana and A.~Loeb,
%``Detecting the earliest galaxies through two new sources of 21cm fluctuations,''
Astrophys. J. \textbf{626}, 1-11 (2005)
doi:10.1086/429954
[arXiv:astro-ph/0410129 [astro-ph]].
%213 citations counted in INSPIRE as of 26 May 2023

%\cite{Barkana:2005jr}
\bibitem{Barkana:2005jr}
R.~Barkana and A.~Loeb,
%``Light-cone anisotropy in 21cm fluctuations during the epoch of reionization,''
Mon. Not. Roy. Astron. Soc. \textbf{372}, L43-L47 (2006)
doi:10.1111/j.1745-3933.2006.00222.x
[arXiv:astro-ph/0512453 [astro-ph]].
%42 citations counted in INSPIRE as of 16 May 2023

%\cite{Mao:2011xp}
\bibitem{Mao:2011xp}
Y.~Mao, P.~R.~Shapiro, G.~Mellema, I.~T.~Iliev, J.~Koda and K.~Ahn,
%``Redshift Space Distortion of the 21cm Background from the Epoch of Reionization I: Methodology Re-examined,''
Mon. Not. Roy. Astron. Soc. \textbf{422}, 926-954 (2012)
doi:10.1111/j.1365-2966.2012.20471.x
[arXiv:1104.2094 [astro-ph.CO]].
%104 citations counted in INSPIRE as of 18 May 2023

%\cite{Hirata:2005mz}
\bibitem{Hirata:2005mz}
C.~M.~Hirata,
%``Wouthuysen-Field coupling strength and application to high-redshift 21 cm radiation,''
Mon. Not. Roy. Astron. Soc. \textbf{367}, 259-274 (2006)
doi:10.1111/j.1365-2966.2005.09949.x
[arXiv:astro-ph/0507102 [astro-ph]].
%207 citations counted in INSPIRE as of 18 May 2023

%\cite{Cai:2004dk}
\bibitem{Cai:2004dk}
R.~G.~Cai and A.~Wang,
%``Cosmology with interaction between phantom dark energy and dark matter and the coincidence problem,''
JCAP \textbf{03}, 002 (2005)
doi:10.1088/1475-7516/2005/03/002
[arXiv:hep-th/0411025 [hep-th]].
%318 citations counted in INSPIRE as of 26 May 2023

%\cite{Amendola:2006dg}
\bibitem{Amendola:2006dg}
L.~Amendola, G.~Camargo Campos and R.~Rosenfeld,
%``Consequences of dark matter-dark energy interaction on cosmological parameters derived from SNIa data,''
Phys. Rev. D \textbf{75}, 083506 (2007)
doi:10.1103/PhysRevD.75.083506
[arXiv:astro-ph/0610806 [astro-ph]].
%228 citations counted in INSPIRE as of 26 May 2023

%\cite{Li:2009zs}
\bibitem{Li:2009zs}
M.~Li, X.~D.~Li, S.~Wang, Y.~Wang and X.~Zhang,
%``Probing interaction and spatial curvature in the holographic dark energy model,''
JCAP \textbf{12}, 014 (2009)
doi:10.1088/1475-7516/2009/12/014
[arXiv:0910.3855 [astro-ph.CO]].
%135 citations counted in INSPIRE as of 26 May 2023

%\cite{Li:2013bya}
\bibitem{Li:2013bya}
Y.~H.~Li and X.~Zhang,
%``Large-scale stable interacting dark energy model: Cosmological perturbations and observational constraints,''
Phys. Rev. D \textbf{89}, no.8, 083009 (2014)
doi:10.1103/PhysRevD.89.083009
[arXiv:1312.6328 [astro-ph.CO]].
%81 citations counted in INSPIRE as of 09 May 2023

%\cite{Geng:2017apd}
\bibitem{Geng:2017apd}
C.~Q.~Geng, C.~C.~Lee and L.~Yin,
%``Constraints on running vacuum model with $H(z)$ and $f \sigma_8$,''
JCAP \textbf{08}, 032 (2017)
doi:10.1088/1475-7516/2017/08/032
[arXiv:1704.02136 [astro-ph.CO]].
%19 citations counted in INSPIRE as of 26 May 2023

%\cite{Zhang:2018nga}
\bibitem{Zhang:2018nga}
J.~J.~Zhang, L.~Yin and C.~Q.~Geng,
%``Cosmological constraints on $\Lambda(\alpha)$CDM models with time-varying fine structure constant,''
Annals Phys. \textbf{397}, 400-409 (2018)
doi:10.1016/j.aop.2018.08.015
[arXiv:1809.04218 [astro-ph.CO]].
%5 citations counted in INSPIRE as of 15 Mar 2023

%\cite{Geng:2020mga}
\bibitem{Geng:2020mga}
C.~Q.~Geng, C.~C.~Lee and L.~Yin,
%``Constraints on a special running vacuum model,''
Eur. Phys. J. C \textbf{80}, no.1, 69 (2020)
doi:10.1140/epjc/s10052-020-7653-z
[arXiv:2001.05092 [astro-ph.CO]].
%10 citations counted in INSPIRE as of 16 May 2023

%\cite{Geng:2020upf}
\bibitem{Geng:2020upf}
C.~Q.~Geng, Y.~T.~Hsu, L.~Yin and K.~Zhang,
%``Running vacuum model in non-flat universe,''
Chin. Phys. C \textbf{44}, 105104 (2020)
doi:10.1088/1674-1137/abab86
[arXiv:2002.05290 [astro-ph.CO]].
%13 citations counted in INSPIRE as of 15 May 2023

%\cite{BOSS:2016wmc}
\bibitem{BOSS:2016wmc}
S.~Alam \textit{et al.} [BOSS],
%``The clustering of galaxies in the completed SDSS-III Baryon Oscillation Spectroscopic Survey: cosmological analysis of the DR12 galaxy sample,''
Mon. Not. Roy. Astron. Soc. \textbf{470}, no.3, 2617-2652 (2017)
doi:10.1093/mnras/stx721
[arXiv:1607.03155 [astro-ph.CO]].
%2044 citations counted in INSPIRE as of 30 May 2023

%\cite{BOSS:2016apd}
\bibitem{BOSS:2016apd}
A.~J.~Ross \textit{et al.} [BOSS],
%``The clustering of galaxies in the completed SDSS-III Baryon Oscillation Spectroscopic Survey: Observational systematics and baryon acoustic oscillations in the correlation function,''
Mon. Not. Roy. Astron. Soc. \textbf{464}, no.1, 1168-1191 (2017)
doi:10.1093/mnras/stw2372
[arXiv:1607.03145 [astro-ph.CO]].
%200 citations counted in INSPIRE as of 16 May 2023

%\cite{Vargas-Magana:2016imr}
\bibitem{Vargas-Magana:2016imr}
M.~Vargas-Maga\~na, S.~Ho, A.~J.~Cuesta, R.~O'Connell, A.~J.~Ross, D.~J.~Eisenstein, W.~J.~Percival, J.~N.~Grieb, A.~G.~S\'anchez and J.~L.~Tinker, \textit{et al.}
%``The clustering of galaxies in the completed SDSS-III Baryon Oscillation Spectroscopic Survey: theoretical systematics and Baryon Acoustic Oscillations in the galaxy correlation function,''
Mon. Not. Roy. Astron. Soc. \textbf{477}, no.1, 1153-1188 (2018)
doi:10.1093/mnras/sty571
[arXiv:1610.03506 [astro-ph.CO]].
%89 citations counted in INSPIRE as of 16 May 2023

%\cite{BOSS:2016hvq}
\bibitem{BOSS:2016hvq}
F.~Beutler \textit{et al.} [BOSS],
%``The clustering of galaxies in the completed SDSS-III Baryon Oscillation Spectroscopic Survey: baryon acoustic oscillations in the Fourier space,''
Mon. Not. Roy. Astron. Soc. \textbf{464}, no.3, 3409-3430 (2017)
doi:10.1093/mnras/stw2373
[arXiv:1607.03149 [astro-ph.CO]].
%196 citations counted in INSPIRE as of 30 May 2023

%\cite{Pan-STARRS1:2017jku}
\bibitem{Pan-STARRS1:2017jku}
D.~M.~Scolnic \textit{et al.} [Pan-STARRS1],
%``The Complete Light-curve Sample of Spectroscopically Confirmed SNe Ia from Pan-STARRS1 and Cosmological Constraints from the Combined Pantheon Sample,''
Astrophys. J. \textbf{859}, no.2, 101 (2018)
doi:10.3847/1538-4357/aab9bb
[arXiv:1710.00845 [astro-ph.CO]].
%1674 citations counted in INSPIRE as of 31 May 2023

%\cite{Riess:2019cxk}
\bibitem{Riess:2019cxk}
A.~G.~Riess, S.~Casertano, W.~Yuan, L.~M.~Macri and D.~Scolnic,
%``Large Magellanic Cloud Cepheid Standards Provide a 1\% Foundation for the Determination of the Hubble Constant and Stronger Evidence for Physics beyond $\Lambda$CDM,''
Astrophys. J. \textbf{876}, no.1, 85 (2019)
doi:10.3847/1538-4357/ab1422
[arXiv:1903.07603 [astro-ph.CO]].
%1513 citations counted in INSPIRE as of 31 May 2023

%\cite{Lesgourgues:2011re}
\bibitem{Lesgourgues:2011re}
J.~Lesgourgues,
%``The Cosmic Linear Anisotropy Solving System (CLASS) I: Overview,''
[arXiv:1104.2932 [astro-ph.IM]].
%793 citations counted in INSPIRE as of 31 May 2023

%\cite{Blas:2011rf}
\bibitem{Blas:2011rf}
D.~Blas, J.~Lesgourgues and T.~Tram,
%``The Cosmic Linear Anisotropy Solving System (CLASS) II: Approximation schemes,''
JCAP \textbf{07}, 034 (2011)
doi:10.1088/1475-7516/2011/07/034
[arXiv:1104.2933 [astro-ph.CO]].
%1485 citations counted in INSPIRE as of 31 May 2023

%\cite{Munoz:2021psm}
\bibitem{Munoz:2021psm}
J.~B.~Mu\~noz, Y.~Qin, A.~Mesinger, S.~G.~Murray, B.~Greig and C.~Mason,
%``The impact of the first galaxies on cosmic dawn and reionization,''
Mon. Not. Roy. Astron. Soc. \textbf{511}, no.3, 3657-3681 (2022)
doi:10.1093/mnras/stac185
[arXiv:2110.13919 [astro-ph.CO]].
%37 citations counted in INSPIRE as of 22 May 2023

%\cite{Pober:2012zz}
\bibitem{Pober:2012zz}
J.~C.~Pober, A.~R.~Parsons, D.~R.~DeBoer, P.~McDonald, M.~McQuinn, J.~E.~Aguirre, Z.~Ali, R.~F.~Bradley, T.~C.~Chang and M.~F.~Morales,
%``The Baryon Acoustic Oscillation Broadband and Broad-beam Array: Design Overview and Sensitivity Forecasts,''
Astron. J. \textbf{145}, 65 (2013)
doi:10.1088/0004-6256/145/3/65
[arXiv:1210.2413 [astro-ph.CO]].
%118 citations counted in INSPIRE as of 15 May 2023

%\cite{Pober:2013jna}
\bibitem{Pober:2013jna}
J.~C.~Pober, A.~Liu, J.~S.~Dillon, J.~E.~Aguirre, J.~D.~Bowman, R.~F.~Bradley, C.~L.~Carilli, D.~R.~DeBoer, J.~N.~Hewitt and D.~C.~Jacobs, \textit{et al.}
%``What Next-Generation 21 cm Power Spectrum Measurements Can Teach Us About the Epoch of Reionization,''
Astrophys. J. \textbf{782}, 66 (2014)
doi:10.1088/0004-637X/782/2/66
[arXiv:1310.7031 [astro-ph.CO]].
%245 citations counted in INSPIRE as of 16 May 2023
	
\end{thebibliography}
\end{document}